\documentclass{article}
\usepackage{ijcai25}

\usepackage{times}
\usepackage{soul}
\usepackage{url}
\usepackage[hidelinks]{hyperref}
\usepackage[utf8]{inputenc}
\usepackage[small]{caption}
\usepackage{graphicx}
\usepackage{amsmath}
\usepackage{amsthm}
\usepackage{booktabs}
\usepackage{algorithm}
\usepackage{algorithmic}
\usepackage[switch]{lineno}
\usepackage{amssymb}
\usepackage{xcolor}

\title{Advancing Community Detection with Graph Convolutional Neural Networks: Bridging Topological and Attributive Cohesion}

\author{
Anjali de Silva$^1$
\and
Gang Chen$^1$\and
Hui Ma$^{1}$\and
Seyed Mohammad Nekooei$^2$\And
Xingquan Zuo$^3$\\
\affiliations
$^1$Victoria University of Wellington, Wellington, New Zealand\\
$^2$Goldenset Collective\\
$^3$School of Computer Science, Beijing University of Posts and Telecommunications, Beijing 100876, China\\
\emails
\{desilanja, aaron.chen, Hui.Ma\}@ecs.vuw.ac.nz,
mohammad@goldenset.com,
zuoxq@bupt.edu.cn
}

\begin{document}

\maketitle

\begin{abstract}
    Community detection, a vital technology for real-world applications, uncovers cohesive node groups (communities) by leveraging both topological and attribute similarities in social networks. However, existing Graph Convolutional Networks (GCNs) trained to maximize modularity often converge to suboptimal solutions. Additionally, directly using human-labeled communities for training can undermine topological cohesiveness by grouping disconnected nodes based solely on node attributes. We address these issues by proposing a novel \emph{Topological and Attributive Similarity-based Community detection} (TAS-Com) method. TAS-Com introduces a novel loss function that exploits the highly effective and scalable Leiden algorithm to detect community structures with global optimal modularity. Leiden is further utilized to refine human-labeled communities to ensure connectivity within each community, enabling TAS-Com to detect community structures with desirable trade-offs between modularity and compliance with human labels. Experimental results on multiple benchmark networks confirm that TAS-Com can significantly outperform several state-of-the-art algorithms. 
\end{abstract}

\section{Introduction}\label{S:Introduction}

\emph{Community Detection} (CD) in social networks is an active research area with immense practical implications across numerous fields like marketing, sociology, and public health \cite{chunaev2020community,wu2020deep}. Unveiling hidden structures within complex networks drives transformative innovations that help to shape the future of society \cite{su2022comprehensive,he2021self}. Real-world applications of CD are vast and impactful, including information spreading \cite{karatacs2018application}, dimensionality reduction, and product recommendation \cite{moradi2015effective}.


Many existing methods for CD are mainly based on network topology \cite{traag2019louvain,de2023leiden,jin2021survey}. However, real-world networks often include node \emph{attributes} that are crucial for identifying meaningful communities. Networks with these attributes are known as \emph{attributed networks} \cite{chunaev2020community,bhowmick2024dgcluster,xie2021effective}.

In an attributed network, a \emph{community} is a group of nodes that are densely connected and share strong attribute similarities \cite{bhowmick2024dgcluster,zhu2024every,luo2020adaptive}. Such communities capture both topological and attribute cohesiveness, making them meaningful representations of real-world groups where both connections and shared characteristics are important. A set of non-overlapping communities that completely covers the whole attributed network is known as the \emph{community structure} of the network. 

CD in attributed networks has been approached through various methods, including heuristic-based \cite{combe2015louvain}, evolutionary computation (EC)-based \cite{guo2024evolutionary}, and learning-based techniques \cite{jin2019graph,bhowmick2024dgcluster,ju2023glcc,he2022semi}. Among these, learning-based methods have gained significant attention in recent literature \cite{sun2019vgraph}, with Graph Convolutional Neural Networks (GCNs) emerging as the leading solution. GCNs leverage deep learning to integrate topological and attribute information through layer-wise propagation \cite{tsitsulin2023graph,he2021community}, offering a precise understanding of network topology and providing a comprehensive view of social networks. Despite their strengths, state-of-the-art GCN approaches still face two critical challenges.

The {\bf first} issue lies in the challenge of directly training GCNs to maximize modularity, which often results in suboptimal community structures \cite{chunaev2020community}. Although maximizing modularity effectively promotes topological cohesiveness \cite{bhowmick2024dgcluster,tsitsulin2023graph}, the inherent complexity of the modularity function makes gradient-based optimization prone to local optima, leading to poor results. This difficulty in achieving globally optimal solutions highlights the limitations of modularity as a direct training objective and underscores the need to use a different loss function that makes it more straightforward to maximize modularity.

%

The \textbf{second} issue arises from using human-labeled communities to train GCNs, as these labels are typically based solely on node attributes while overlooking topological cohesiveness. This omission, a critical factor for high-quality community structures, often leads to poorly formed communities with disconnected nodes, significantly reducing modularity. This problem highlights the necessity of integrating topological information with node attributes to enhance GCNs' ability to produce well-formed and cohesive community structures. 





%

To address these issues, we propose a novel \emph{Topological and Attributive Similarity-based Community detection} (TAS-Com\footnote{The code is available at \url{https://github.com/desilanja/TAS-Com}}). TAS-Com introduces a newly designed loss function to guide GCN training. Our innovative loss function exploits the highly effective and scalable Leiden algorithm ~\cite{traag2019louvain} (see Appendix M for more details) to identify community structures with global optimal modularity. Leiden is further utilized to refine human-labeled communities to ensure connectivity within each community, enabling TAS-Com to detect community structures with desirable trade-offs between modularity and compliance with human labels. The key contributions of this paper are as follows:
\begin{itemize}
    \item We are the first to address the limitations of directly using human-provided community labels to train GCNs, which often compromise CD due to lack of topological cohesiveness. To overcome this, we propose a novel Leiden-based method to refine human-labeled communities, ensuring they are both cohesive and free from disconnected nodes that degrade CD quality, thereby enhancing their suitability for effective GCN training.
    

    \item We are the first to address the limitations of directly training GCNs to maximize modularity. We propose a novel \emph{modularity-based similarity loss}, enhanced by the Leiden algorithm, to effectively guide GCNs toward discovering community structures with globally optimal modularity. Additionally, we integrate this loss with a \emph{refined human-label-based similarity loss}, significantly improving the GCN’s ability to identify community structures that achieve both high modularity and strong alignment with human-labeled communities. 
    
    \item We conduct extensive experiments on many real-world attributed benchmark networks. The experiment results confirm that TAS-Com can significantly outperform multiple state-of-the-art approaches for CD across most of these benchmark networks.
\end{itemize}

\section{Related Work}\label{S:RelatedNetworks}


CD has traditionally focused on graph topology, but recent research \cite{chunaev2020community} started to consider node attributes to improve CD in attributed networks. Heuristic approaches like I-Louvain \cite{combe2015louvain} use inertia-based measures for attribute similarity but often fail to find globally optimal structures. EC methods, such as @NetGA \cite{pizzuti2018genetic}, optimize fitness functions combining attributes and connectivity, while multi-objective algorithms \cite{li2017multiobjective} refine these strategies. However, these approaches rely on manually designed similarity measures, overlooking intrinsic node relationships and highlighting the need for machine learning techniques to learn high-level node embeddings for more accurate CD \cite{su2022comprehensive}.



The emergence of advanced Graph Neural Networks (GNNs) has significantly shifted recent research toward learning-based approaches, showcasing the critical importance of leveraging these technologies for more effective CD \cite{chunaev2020community,zhu2024every,yang2023cluster,liu2022deep,xia2021self}. These approaches generally perform CD through two consecutive steps: 1) node representation learning, and 2) node grouping/clustering based on the learned high-level node embeddings \cite{zhou2023community,tsitsulin2023graph,bhowmick2024dgcluster}.

The most commonly used GNNs for CD are Autoencoders (AEs) \cite{zhu2024every} and GCNs \cite{bhowmick2024dgcluster}. AEs are proficient at extracting node semantic information \cite{liu2024information,sun2020network,kumar2023community}. Notable AE-based approaches include DNR \cite{yang2016modularity} and CDBNE \cite{zhou2023community}. Despite their recent success, AEs are not specifically designed to process graph data. In contrast, GCNs can effectively handle node information at both topological and attribute levels~\cite{zhu2024every}.

Most GCN approaches for CD conduct either supervised or semi-supervised learning based on ground truth community labels \cite{bhowmick2024dgcluster}. A few recent studies, like SGCN \cite{wang2021unsupervised} and DMoN \cite{tsitsulin2023graph}, further demonstrated the importance of unsupervised learning. For example, Zhu et al. \cite{zhu2024every} introduced DyFSS, an unsupervised approach that dynamically fuses embeddings from multiple self-supervised tasks with node-specific weights, effectively balancing attribute and structural information to enhance clustering accuracy and robustness. However, DyFSS’s reliance on hyperparameter tuning, such as pseudo-label thresholds, limits its generalizability across diverse graph datasets without extensive adjustments.

Existing unsupervised methods largely focus on optimizing modularity, often neglecting explicit optimization of attribute similarity \cite{tsitsulin2023graph}. While DGCluster \cite{bhowmick2024dgcluster} optimizes both modularity and attribute similarity, its maximization of modularity frequently leads to suboptimal community structures. DGCluster learns directly from human-labeled communities, which often group disconnected nodes based solely on attributes, undermining connectivity cohesiveness. To overcome these challenges, we propose TAS-Com, powered by an innovative loss function that properly integrates topological and attribute information, improves the quality of human-labeled communities, and enables GCNs to discover community structures with optimal modularity and strong alignment with human labels.



\section{Problem Definition}\label{S:ProblemDefinition}

An \emph{attributed network} can be modeled as a graph $N = (V, E, X)$, where $V$ is the set of nodes, i.e., $V = \{v_1, v_2,..., v_n\}$. $E$ is the set of edges, i.e., $E = \{e_{i,j} | e_{i,j} \in V \times V\}$. $X \in \mathbb{R}^{n \times T}$ is the node attributes matrix, where $T$ is the number of attributes for each node.

A \emph{community structure} $CS$ of an attributed network $N$ is a set of non-overlapping communities, i.e., $CS = \{C_1, C_2, ..., C_k\}$ where $k \geq 1$ s.t. $\forall q \neq l$, $C_q \cap C_l = \emptyset$ and $\cup_{q=1}^{p}C_q = V$. The main goal of CD in attributed network $N$ is to identify $CS$ that satisfies the following conditions:
\begin{itemize}
    \item High topological cohesiveness: This implies that intra-community edge density within any community is higher than the inter-community edge density among different communities \cite{chunaev2020community}. 
    \item High attribute similarity: Nodes within the same community exhibit strong attribute level similarity. Meanwhile, nodes from different communities are expected to have distinct and dissimilar attributes.
\end{itemize}

\emph{Modularity ($Q$)} is a well-known metric to quantify the quality of different community structures \cite{newman2004finding}. As a quality metric at the topology level, $Q$ with respect to any given community structure $CS$ is defined in Equation \eqref{equ:modularity}:
\begin{equation}\label{equ:modularity}
    Q(CS) = \frac{1}{2m}\sum_{ij}(A_{ij}-\frac{k_ik_j}{2m})\delta(c_i,c_j,CS),
\end{equation}
where $m$ is the total number of edges of the social network, $k_i$ and $k_j$ are the degrees of the nodes $v_i$ and $v_j$. $A_{ij}$ is the adjacency matrix where $A_{ij} = 1$ if an edge exits between nodes $v_i$ and $v_j$, otherwise $0$. $\delta(.)$ is the Kronecker delta function where $\delta(c_i,c_j,CS) = 1$ if nodes $v_i$ and $v_j$ are in the same community (i.e., $c_i = c_j$ with $c_i$ and $c_j$ being the community labels of node $i$ and $j$ respectively), otherwise $0$.

\emph{Normalized Mutual Information ($N\!M\!I$)} is a widely used metric in existing research to assess how well a community structure $C$ aligns with human-labeled community structure $D$ in a social network $N$, particularly in terms of attributive similarity \cite{bhowmick2024dgcluster,chunaev2020community}. It is defined in Equation \eqref{equ-nmi} below:
\begin{equation}
\begin{split}
    & N\!M\!I(C,D)  = \\ 
 & \frac{-2\sum_{i=1}^{g_C}\sum_{j=1}^{g_D}P_{ij}\log(P_{ij}n/P_{i.}P_{j.})}{\sum_{i=1}^{g_C}P_{i.}\log(P_{i.}/n)+\sum_{j=1}^{g_D}P_{.j}\log(P_{.j}/n)},
\end{split}
\label{equ-nmi}
\end{equation}
where $P$ represents the confusion matrix. Each of its elements $P_{ij}$ refers to the number of nodes of community $C_i \in C$ that are also in community $D_j \in D$. The values of $i$ and $j$ span within the range of $\{1, \ldots, n\}$ where $n$ is the number of nodes in $N$. $g_C$ refers to the number of communities in $C$. $g_D$ refers to the number of communities in $D$. $P_{i.}$ denotes the sum of the $i$-th row of $P$ and $P_{.j}$ denotes the sum of the $j$-th column. 

\textbf{Our Goal:} We aim to identify high-quality community structures that can outperform those discovered by other state-of-the-art methods, achieving significantly higher scores in both modularity (indicating topological cohesiveness) and $N\!M\!I$ (indicating alignment with human labels).

\section{Proposed Method}\label{S:PrposedMethod}

\begin{figure*}[h!tbp]
\centering
\includegraphics[width=1.0\textwidth]{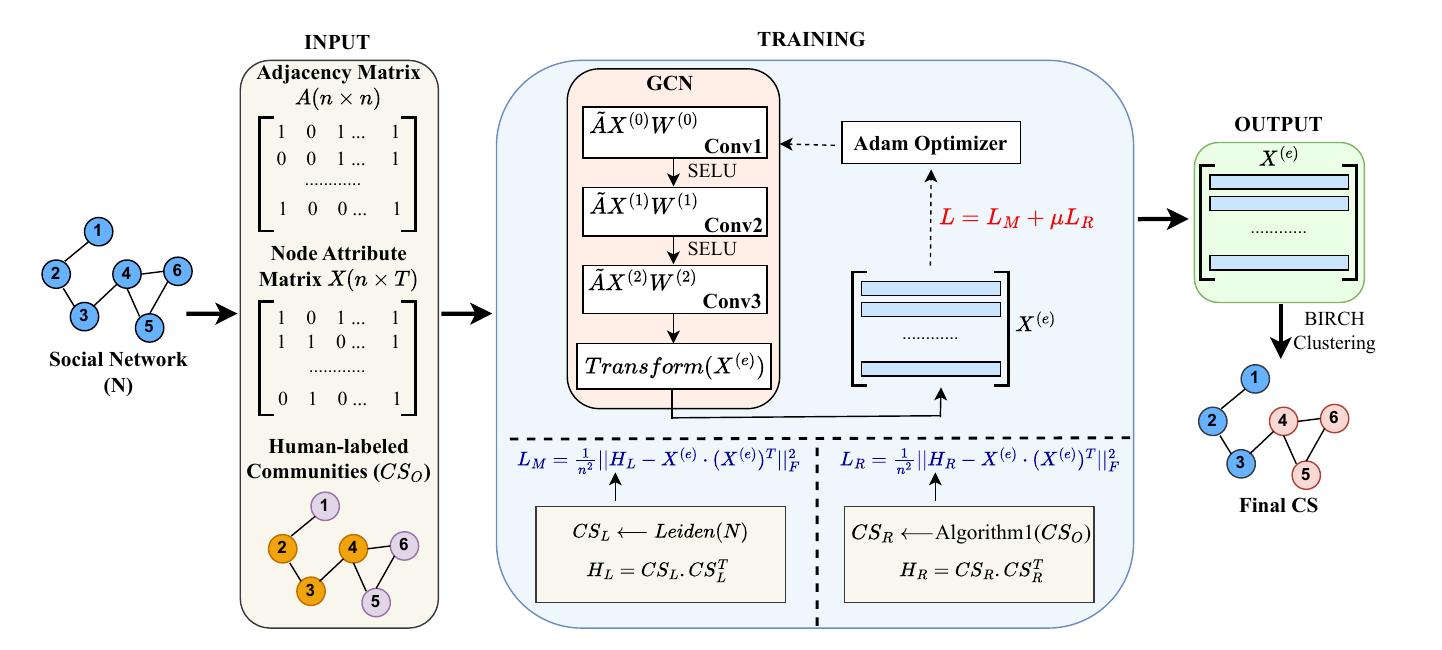} 
\caption{The overall design of the proposed TAS-Com approach.}
\label{fig1:DesignOverview}
\end{figure*}

This section proposes TAS-Com for CD in attributed networks. Figure \ref{fig1:DesignOverview} illustrates the overall design of TAS-Com. As shown in this figure, the adjacency matrix $A$, node attribute matrix $X$, and human-labeled communities $CS_O$ are utilized to build the loss function $L$ and to train our GCN model. Specifically, the GCN model first processes $A$ and $X$ to generate the high-level node embedding $X^{(e)}$. Supported by the Leiden algorithm and our refinement algorithm in Algorithm \ref{alg:refinement}, $X^{(e)}$ is further adopted to calculate respectively the \emph{modularity-based loss} $L_M$ and \emph{refined human-label-based loss} $L_R$ to be introduced later in this section. The total loss $L$ is then constructed by combining $L_M$ and $L_R$. $L$ subsequently guides the training of the GCN. Finally, based on $X^{(e)}$ produced by the trained GCN, a clustering algorithm named BIRCH \cite{zhang1996birch} is performed to obtain the final community structure $CS$. 

We follow \cite{bhowmick2024dgcluster} to design the architecture of our GCN model. In particular, the message passing rule for the $l$-th layer, where $l = {0, 1, ..., L-1}$, is defined in Equation \eqref{equ:NodeEmbedding}:
\begin{equation}\label{equ:NodeEmbedding}
    X^{(l)} = \sigma(\Tilde{A}X^{(l-1)}W^{(l-1)}),
\end{equation}
where $\Tilde{A}$ is the normalized adjacency matrix s.t. $\Tilde{A} = D^{-\frac{1}{2}}AD^{\frac{1}{2}}$. $D$ is the diagonal node degree matrix \cite{bhowmick2024dgcluster}. The embedding output of the $l$-th layer is denoted by $X^{(l)}$. $W^{(l)}$ refers to the corresponding learnable weight matrix of this layer. SELU is used as the activation function $\sigma(\cdot)$ to incorporate the non-linearity for the aggregation of node attributes \cite{klambauer2017self}. Further, the node embedding matrix $X^{(e)}$ is transformed as in \cite{bhowmick2024dgcluster} to ensure that the embedding is constrained within the positive coordinate space (see Appendix C). The following provides detailed descriptions of the new components introduced in TAS-Com.

\subsection{Proposed loss function ($L$)}\label{LossFunction}
In line with the problem formulation in the previous section, we propose a new loss function to train GCNs to jointly maximize the node connectivity strength (modularity optimization) and the node attribute similarity ($N\!M\!I$ optimization). The proposed loss function is defined at a high level in Equation \eqref{equ:L}:
\begin{equation}\label{equ:L}
    L = L_M + \mu L_R,
\end{equation}
where $\mu\geq 0$ is a hyperparameter that controls the influence of $L_R$ in $L$. $L_M$ refers to the \emph{modularity-based similarity loss} and $L_R$ refers to the \emph{refined human-label-based similarity loss}. The process of constructing $L_M$ and $L_R$ are explained in the next two subsections.  

\subsubsection{Modularity-based similarity loss ($L_M$)}
Most existing GNNs were trained directly to optimize the modularity metric (see Equation \eqref{equ:modularity}), resulting frequently in locally optimal community structures with low modularity \cite{tsitsulin2023graph}. In this paper, we employ the Leiden algorithm with proven effectiveness and scalability \cite{traag2019louvain,de2023leiden,de2022genetic} to identify high-quality community structures with close-to-optimal modularity. Due to the stochastic nature of Leiden, we apply Leiden to a social network 30 times and select the community structure with the highest $N\!M\!I$ score out of all runs. Leiden focuses solely on topological information, whereas our problem also concerns about attribute information. Therefore, we select the community structure with the highest NMI from 30 runs, ensuring an ideal balance between topological level and attribute level similarities.

Let $H \in \mathbb{R}^{n \times n}$ be the pairwise information matrix \cite{bhowmick2024dgcluster}. As an $n\times n$ matrix, where $n$ is the total number of nodes in a social network, each element of $H$ is defined as in Equation \eqref{equ:H}:
\begin{equation}\label{equ:H}
    H_{ij} = 
    \begin{cases}
      1 & \text{if $c_i = c_j$};\\
      0 & \text{otherwise},
    \end{cases}
\end{equation}
where $c_i$ and $c_j$ are the community assignments of nodes $v_i$ and $v_j$ according to Leiden. Specifically, the community structure identified by Leiden can be represented in the form of a one-hot matrix $CS_L \in \{0,1\}^{n \times k}$ below:
$$
(CS_L)_{ij} = 
\begin{cases}
      1 & \text{if node $v_i$ belongs to community $j$};\\
      0 & \text{otherwise},
    \end{cases}
$$
where $k$ is the total number of communities detected by Leiden. Subsequently, $H$ can be rewritten as follows:
\begin{equation}\label{equ:newHL}
    H_L = CS_L\cdot CS_L^T.
\end{equation}
In line with the above, the modularity-based loss $L_M$ is designed to minimize the discrepancy between the community assignments obtained by Leiden and the embedding similarity matrix derived from the high-level node embedding $X^{(e)}$ produced by the GCN model, as detailed below:
\begin{equation}\label{equ:LM}
    L_M = \frac{1}{n^2}||H_L - X^{(e)}\cdot (X^{(e)})^T||_F^2.
\end{equation}

\subsubsection{Refined human-label-based similarity loss ($L_R$)}\label{SS:Refinement}
As pointed out in the introduction, human-labeled communities contain disconnected nodes that seriously hurt the quality of the community structure in terms of modularity (see Appendix D for a concrete example). Hence, we propose a new method to refine human-labeled communities to simultaneously enhance consistency with human labels and connectivity within each community. In practice, if human-labeled communities are unavailable, any scalable CD algorithm can be used to obtain the substitute labels. The pseudo-code of the newly proposed refinement method is presented in Algorithm \ref{alg:refinement}. According to this algorithm, the refinement process consists of two main steps explained below.



\begin{algorithm}[tb]
\caption{Refinement of the human-labeled communities}
\label{alg:refinement}
\textbf{Input}: Human-labeled community structure $CS_O$\\
\textbf{Output}: Refined community structure $CS_R$

\begin{algorithmic}[1] 
\FOR{each community $C_k$ in $CS_O$}
    \STATE \textbf{Step 1: Apply Leiden algorithm}
    \STATE Obtain a sub-network $SN$ that contains nodes in $C_k$
    \STATE $CS_L \longleftarrow CS $ with highest $Q(CS_L)$ after applying $Leiden(SN)$ for $n$ times
    \STATE \textbf{Step 2: Merge sub-communities}
    \WHILE{$|CS_L| \geq threshold$}
        \FOR{each pair of sub-communities $C_i$ and $C_j$ in $CS_L$}
            \STATE $CS_{new} \longleftarrow$ Merge $C_i$ and $C_j$
            \STATE Calculate $Q(CS_{new})$ \COMMENT{Refer Equation \eqref{equ:modularity}}
        \ENDFOR
        \STATE $CS_L \longleftarrow $ $CS_{new}$ with the highest $Q(CS_{new})$
    \ENDWHILE
    \STATE $CS_R \longleftarrow  (CS_R\setminus \{C_k\}) \cup CS_L $
\ENDFOR
\STATE \textbf{return} $CS_R$
\end{algorithmic}
\end{algorithm}

\begin{enumerate}
    \item \textbf{Apply a topology-based CD algorithm to detect sub-communities in each human-labeled community.} Since Leiden can produce topologically connected communities~\cite{traag2019louvain}, we apply Leiden to obtain the connected sub-communities within each human-labeled community. In practice, any scalable CD algorithm, such as Louvain \cite{blondel2008fast}, can replace Leiden for this task. Due to the stochastic nature of Leiden, it is executed 10 times in our experiments for each human-labeled community. We then choose the community structure with the highest modularity, i.e., $CS_L$, as described in lines 2--4 of Algorithm \ref{alg:refinement}.
    
    \item \textbf{Merge sub-communities.} In our experiments (see Appendix E), we found that Leiden can sometimes produce many small sub-communities $CS_L$ that significantly deviate from human-labeled communities. Hence, GCNs trained based on $CS_L$ can exhibit poor $N\!M\!I$ performance. To address this issue, we must merge closely connected sub-communities in $CS_L$. For this purpose, we iteratively merge pairs of sub-communities in $CS_L$. Each iteration will select and merge the pair of sub-communities that can produce the highest modularity after merging among all possible merge pairs (see lines 6--12 in Algorithm \ref{alg:refinement}). This process continues until a specific $threshold=\frac{|CC(N_L)|}{2}$ is reached (see Appendix B), where $CC(N_L) = \{C_1, \ldots, C_k\}$ refers to the set of connected components of the sub-network $N_L$ containing all nodes of network $N$ in $CS_L$. 
\end{enumerate}

The community structure obtained from the above two steps is deemed the \emph{refined human-label-based community structure}, denoted as $CS_R$. $CS_R$ is subsequently adopted to train our GCN to maximize $N\!M\!I$. Notably, $CS_R$ is consistent with human-labeled communities since each human-labeled community comprises one or more connected sub-communities in $CS_R$.

Using $CS_R$, we define the refined human-label-based loss $L_R$ in Equation \eqref{equ:LR} below:
\begin{equation}\label{equ:LR}
    L_R = \frac{1}{n^2}||H_R - X^{(e)}\cdot (X^{(e)})^T||_F^2.
\end{equation}
In the above, matrix $H_R$ follows the definition in Equation \eqref{equ:H}. It is the pairwise matrix obtained from below:
\begin{equation}\label{equ:HR}
    H_R = CS_R\cdot CS_R^T,
\end{equation}
where $CS_R$ is the one-hot matrix (similar to $CS_L$) derived from the community structure obtained from Algorithm \ref{alg:refinement}. According to Equation \eqref{equ:LR}, $L_R$ guides our GCN to minimize the discrepancy between $H_R$ and the embedding similarity matrix derived from $X^{(e)}$. This is expected to enable the trained GCN to produce node embedding that closely aligns with human-provided community labels.

Driven by the loss function $L$ designed above, we can finally develop a system to train GCNs through the stochastic gradient descent method. The training process is summarized in Appendix A.

\subsection{Community detection based on node embedding}
Based on the node embedding produced by the trained GCN model, we utilize the clustering algorithm named Balance Iterative Reducing and Clustering using Hierarchies (BIRCH) \cite{zhang1996birch} to identify the final community structure, following DGCluster \cite{bhowmick2024dgcluster}. The primary benefit of BIRCH is its ability to flexibly construct community structures without determining the number of communities in advance, enabling the algorithm to achieve a desirable trade-off between modularity and $N\!M\!I$. 

\section{Experiment Design}\label{S:ExperimentDesign}
This section outlines the experimental settings used to examine the performance of TAS-Com.

\subsection{Benchmark networks}\label{SS:BenchmarkNetworks}
We conduct experiments on six commonly used attributed social networks with human-provided community labels ~\cite{bhowmick2024dgcluster,zhu2024every}. In particular, Cora and Citeseer are citation networks \cite{sen2008collective}, Amazon Photo and Amazon PC are co-purchase networks \cite{shchur2018pitfalls}, and Coauthor CS \cite{shchur2018pitfalls} and Coauthor Phy \cite{shchur2019overlapping} are co-authorship networks for computer science and physics respectively. Table \ref{tab:benchmarknetworks} summarizes all the benchmark networks.

\begin{table}[h]
   \centering
   \begin{tabular}{ccccc}
   \toprule
   \textbf{Network} & \textbf{$n$} & \textbf{$m$} & \textbf{$T$} & \textbf{$k$}\\ 
   \midrule
    Cora & 2708 & 5278 & 1433 & 7\\
    Citeseer & 3327 & 4552 & 3703 & 6\\
    Amazon Photo & 7650 & 119081 & 745 & 8\\
    Amazon PC & 13752 & 245861 & 767 & 10\\
    Coauthor CS & 18333 & 81894 & 6805 & 15\\
    Coauthor Phy & 34493 & 247962 & 8415 & 5\\
   \bottomrule
   \end{tabular}
   \caption{Statistics of benchmark social networks. $n$, $m$, $T$, and $k$ denote the number of nodes, edges, node attributes, and human-provided community labels respectively.}\label{tab:benchmarknetworks}
\end{table}

\subsection{Baseline approaches}\label{SS:BaselineAlgorithms}

The effectiveness of TAS-Com is evaluated in comparison to 12 state-of-the-art approaches, categorized into three distinct groups: (1) approaches relying solely on attributive or topological information for CD, including k-m(feat) (i.e., k-means based only on features) and DMoN \cite{tsitsulin2023graph}; (2) approaches that consider node similarities at both the topological and attribute levels, including k-m(DW) \cite{perozzi2014deepwalk}, k-means(DGI) \cite{velivckovic2018deep}, DAEGC \cite{wang2019attributed}, SDCN \cite{bo2020structural}, NOCD \cite{shchur2019overlapping}, DyFSS \cite{zhu2024every}, and DGCluster \cite{bhowmick2024dgcluster}, and (3) approaches that use graph pooling techniques, including DiffPool \cite{ying2018graph}, MinCutPool \cite{bianchi2020spectral}, and Ortho \cite{bianchi2020spectral}. Additional details regarding these approaches are provided in Appendix G, while the performance of the MinCutPool and Ortho approaches is analyzed in Appendix J. Among all competing approaches, DGCluster is the most competitive approach, as it is explicitly designed to maximize both modularity and attribute-level node similarity.

\subsection{Performance metrics}\label{SS:PerformanceMetrics}

We adopt two metrics, namely modularity and $N\!M\!I$ introduced in Section \ref{S:ProblemDefinition} to compare the performance of all competing approaches. As explained previously, both metrics are essential to determine the quality of the community structures identified by the trained GCNs~\cite{bhowmick2024dgcluster}. 



\subsection{Parameter settings}\label{ParameterSettings}

In the experiments, TAS-Com employs a GCN architecture identical to that proposed in \cite{bhowmick2024dgcluster}. The model uses a GCN with two hidden layers, the Adam optimizer with a learning rate of 0.001, and is trained for 300 epochs. Additional details on the parameter settings of TAS-Com are provided in Appendix K. In line with existing works \cite{bhowmick2024dgcluster,tsitsulin2023graph}, we report the average performance results across 10 independent runs with different random seeds. Furthermore, $\mu$ and $threshold$ in Algorithm \ref{alg:refinement} are two important hyperparameters. Their sensitivity analysis is reported in Appendix B.


\section{Results and Discussion}\label{SS:Results}

\subsection{Performance comparison}

Table \ref{tab:performance} compares the performance of TAS-Com with 12 state-of-the-art approaches in terms of modularity ($Q$) and $N\!M\!I$. Most of the results, except DyFSS and DGCluster have been reported previously in \cite{bhowmick2024dgcluster}. The results for DyFSS and DGCluster were reproduced using their published code and original experimental settings. Furthermore, the results of DGCluster are reported in Table \ref{tab:performance} based on two settings of the hyperparameter $\lambda$ (i.e., $\lambda = \{0.2, 0.8\}$), which are the recommended settings in \cite{bhowmick2024dgcluster}. As evidenced in Table \ref{tab:performance}, TAS-Com achieved the best trade-off between $Q$ and $N\!M\!I$ across all benchmark networks. We have conducted the Wilcoxin rank-sum test. Our test results show that the observed performance gain of TAS-Com is statistically significant (see Appendix H for statistical analysis).

Specifically, while TAS-Com obtained identical $N\!M\!I$ value (i.e., $41.0$) on the Citeseer network as DGCluster, TAS-Com significantly improves the quality of the community structure in terms of modularity $Q$. Although DyFSS attained the highest $N\!M\!I$ value for the Citeseer network, its corresponding $Q$ value was notably lower than that of TAS-Com. Meanwhile, concerning the Amazon PC and Coauthor Phy networks, TAS-Com noticeably improves the quality of the community structures in terms of $N\!M\!I$ (i.e., $12\%$ and $11\%$ increase in $N\!M\!I$ compared to DGCluster on the two networks respectively) without hurting the modularity $Q$. For the Coauthor CS network, while TAS-Com doesn't match the best $Q$ value of $74.2$ achieved by DGCluster($\lambda = 0.2$), it improves the $Q$ value by $1.1\%$ and matches the $N\!M\!I$ value of $82.1$ obtained by DGCluster($\lambda = 0.8$). 

These results confirm that TAS-Com outperforms in $Q$ and/or $N\!M\!I$ without sacrificing either metric. It shows significant performance gains for several benchmark networks, reinforcing its competitiveness for CD in attributed networks.


\begin{table*}[h]
   \centering
   \resizebox{.95\textwidth}{!}{
   \begin{tabular}{ccccccccccccc}
   \toprule
   Approach & \multicolumn{2}{c}{Cora} & \multicolumn{2}{c}{Citeseer} & \multicolumn{2}{c}{Amazon Photo} & \multicolumn{2}{c}{Amazon PC} & \multicolumn{2}{c}{Coauthor CS} & \multicolumn{2}{c}{Coauthor Phy}\\ 
   \midrule
    & $Q$ & $N\!M\!I$ & $Q$ & $N\!M\!I$ & $Q$ & $N\!M\!I$ & $Q$ & $N\!M\!I$ & $Q$ & $N\!M\!I$ & $Q$ & $N\!M\!I$ \\
    \midrule
    k-m(feat) & 19.8 & 18.5 & 30.3 & 24.5 & 10.5 & 28.8 & 5.4 & 21.1 & 23.1 & 35.7 & 19.4 & 30.6 \\ 
    \midrule
    k-m(DW) & 30.7 & 24.3 & 24.3 & 27.6 & 22.9 & 49.4 & 11.8 & 38.2 & 59.4 & 72.7 & 47.0 & 43.5 \\
   \midrule
    SDCN & 50.8 & 27.9 & 62.3 &  31.4 & 53.3 & 41.7 & 45.6 & 24.9 & 55.7 & 59.3 & 52.8 & 50.4\\
   \midrule
   DAEGC & 33.5 & 8.3 & 36.4 & 4.3 & 58.0 & 47.6 & 43.3 & 42.5 & 49.1 & 36.3 & N/A & N/A \\
   \midrule
   k-m(DGI) & 64.0 & 52.7 & 73.7 & 40.4 & 35.1 & 33.4 & 22.8 & 22.6 & 57.8 & 64.6 & 51.2 & 51.0 \\
   \midrule
   NOCD & 78.3 & 46.3 & 84.4 & 20.0 & 70.1 & 62.3 & 59.0 & 44.8 & 72.2 & 70.5 & 65.5 & 28.7 \\
   \midrule
   DiffPool & 66.3 & 32.9 & 63.4 & 20.0 & 46.8 & 35.9 & 30.4 & 22.1 & 59.3 & 41.6 & N/A & N/A \\
   \midrule
   DMoN & 76.5 & 48.8 & 79.3 & 33.7 & 70.1 & 63.3 & 59.0 & 49.3 & 72.4 & 69.1 & 65.8 & 56.7 \\
   \midrule
    DyFSS & 73.5 & 55.5 & 75.2 & 44.8 & 57.7 & 53.6 & 35.1 & 36.3 & 68.9 & 76.8
    & 65.6 & 56.9 \\
    \midrule
   DGCluster($\lambda = 0.2$) & 80.8 & 53.0 & 87.4 & 30.3 & 71.6 & 73.0 & 61.5 & 53.8 & 74.2 & 76.1 & 67.3 & 59.0 \\
   DGCluster($\lambda = 0.8$) & 78.6 & 62.1 & 86.3 & 41.0 & 71.6 & 77.3 & 60.3 & 60.4 & 73.3 & 82.1 & 66.0 & 65.7\\
   \midrule
   \textbf{TAS-Com} & \textbf{81.7} & \textbf{65.1} & \textbf{88.1} & 41.0 & \textbf{72.2} & \textbf{78.4} & \textbf{61.5} & \textbf{61.2} & 74.1 & \textbf{82.1} & \textbf{67.3} & \textbf{66.0} \\
   \bottomrule
   \end{tabular}}
   \caption{Performance comparison in terms of $Q$ and $N\!M\!I$ (results are multiplied by 100) between TAS-Com and state-of-the-art approaches. The bolded results indicate instances where TAS-Com achieves the best results. N/A: Not Available. }\label{tab:performance}
\end{table*}

\subsection{Further analysis}
We conduct additional analysis to validate the effectiveness of TAS-Com.

\subsubsection{Connectivity within communities}

\begin{figure}[h]
\centering
\includegraphics[width=1.0\columnwidth]{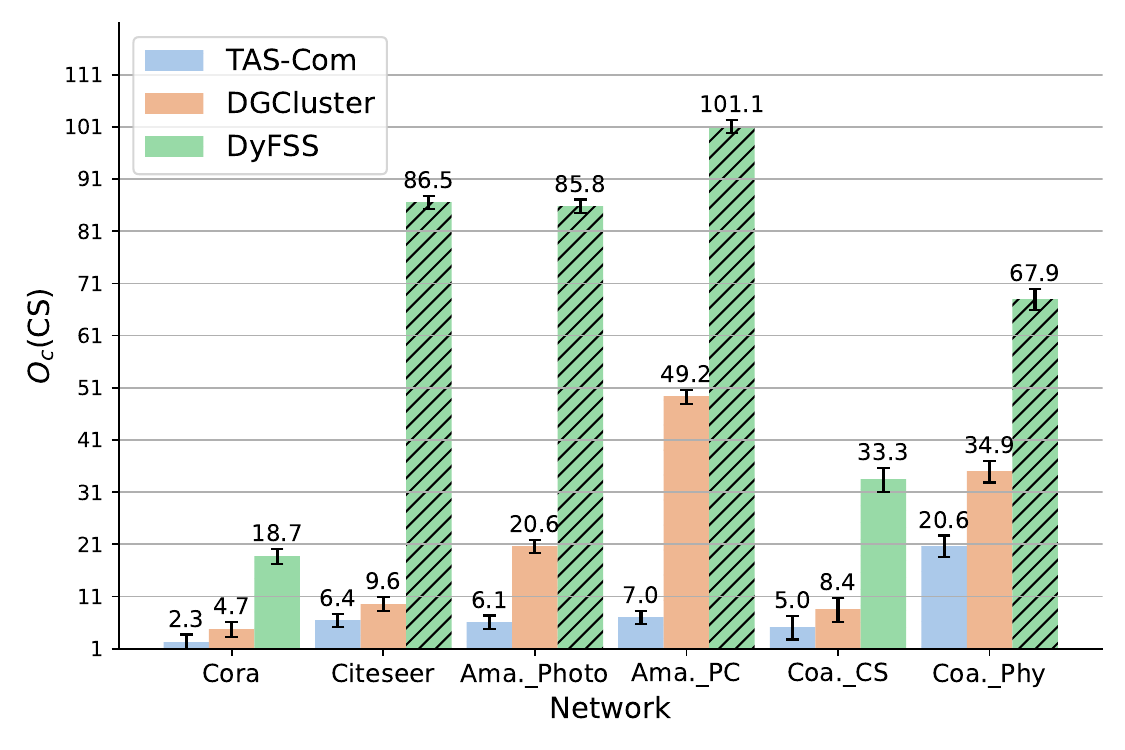} 
\caption{Comparison of $O_c(CS)$ achieved by TAS-Com, DGCluster, and DyFSS across all benchmark networks.}
\label{fig2:CommunityConnectivity}
\end{figure}

We analyze the connectivity within each community identified by TAS-Com, DGCluster, and DyFSS. For this purpose, we check the average number of isolated sub-networks in each community of a community structure $CS = \{C_1, C_2, \ldots, C_k\}$, i.e., $O_c(CS)$, as defined below: 

\begin{equation}\label{equ:connectivity}
    O_c(CS) = \frac{\sum_{C\in CS} |CC(C)|}{|CS|},
\end{equation}
where $CC(C)$ refers to the set of isolated sub-networks within community $C\in CS$. Ideally, all nodes with any given community $C$ are expected to be inter-connected, i.e., $|CC(C)| = 1$. In this case, $O_c(CS)$ reaches its smallest value of $1$.   
Hence, the lower the value of $O_c(CS)$, the better. Figure \ref{fig2:CommunityConnectivity} compares $O_c(CS)$ achieved respectively by TAS-Com, DGCluster, and DyFSS. For DGCluster, we report the lowest $O_c(CS)$ obtained by either DGCluster($\lambda = 0.2$) or DGCluster($\lambda = 0.8$). 

Figure \ref{fig2:CommunityConnectivity} shows that TAS-Com significantly reduces isolated sub-networks compared to DGCluster and DyFSS.  While TAS-Com may still produce isolated sub-networks, these can be treated as separate communities in the final $CS$ without reducing modularity $Q$ (see Appendix F). This confirms that using refined human-labeled communities to guide the training of GCNs enhances the effectiveness of CD.

\subsubsection{Performance evaluation using additional metrics}

To further evaluate the performance of the proposed TAS-Com, we employ two additional metrics: \emph{conductance} and \emph{F1 score} \cite{bhowmick2024dgcluster}(see Appendix L). Table \ref{tab:performanceConF1} compares TAS-Com and the baseline algorithm, DGCluster, in terms of average conductance ($Con$) and average F1 score ($F1$) on the benchmark networks Cora and Amazon Photo. The results demonstrate that TAS-Com outperforms DGCluster on these networks, highlighting its superior performance not only in $Q$ and $N\!M\!I$ but also in $Con$ and $F1$. Additional results for all benchmark networks are in Appendix I.

\begin{table}[h]
   \centering
   \begin{tabular}{ccccc}
   \toprule
   Approach & \multicolumn{2}{c}{Cora} & \multicolumn{2}{c}{Amazon Photo} \\ 
   \midrule
    & $Con$ & $F1$ & $Con$ & $F1$ \\
    \midrule
   DGCluster($\lambda = 0.2$) & 9.7 & 43.5 & 8.6 & 70.7 \\
   DGCluster($\lambda = 0.8$) & 14.5 & 54.5 & 12.4 & 75.9 \\
   \midrule
   \textbf{TAS-Com} & \textbf{8.8} & \textbf{56.9} & \textbf{8.1} & \textbf{76.7} \\
   \bottomrule
   \end{tabular}
   \caption{Performance comparison of TAS-Com and DGCluster in terms of $Con$ (the lower the better) and $F1$ (the higher the better). The best results (multiplied by 100) are bolded.}\label{tab:performanceConF1}
\end{table}

\subsubsection{Ablation study }

\begin{figure}[h]
\centering
\includegraphics[width=0.8\columnwidth]{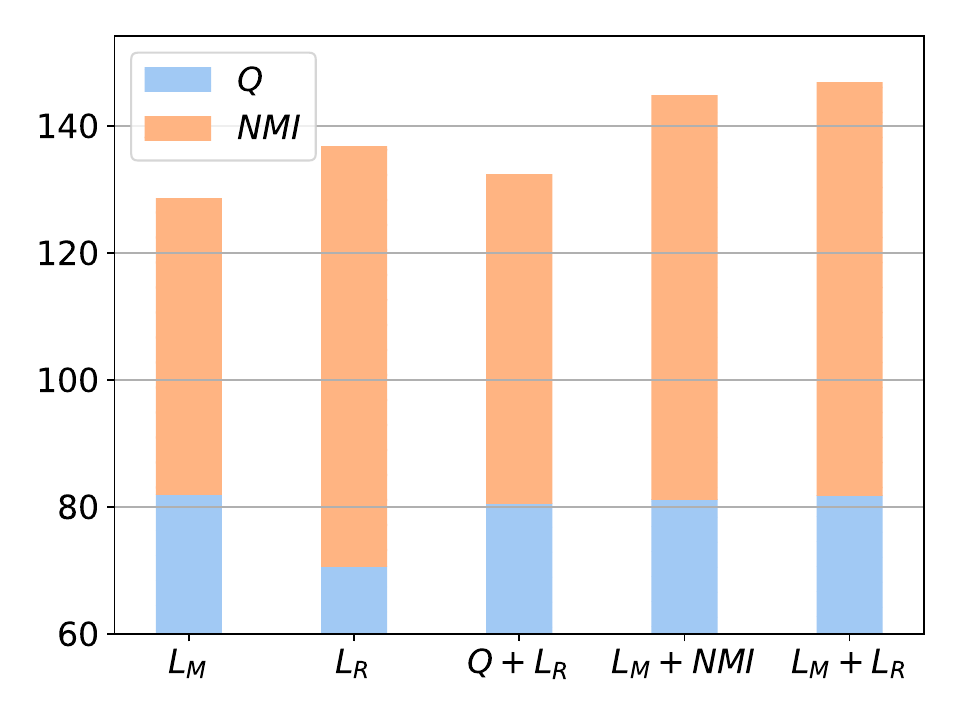} 
\caption{Ablation study on the Cora network evaluating the impact of $L_M$ and $L_R$ in the loss function $L$. Scores ($Q$ and $N\!M\!I$) are scaled by 100. $L_M+L_R$ achieves the best total score.}
\label{fig2:AblationStudy}
\end{figure}

To verify the effectiveness of the loss function proposed in Equation~\eqref{equ:L}, an ablation study has been performed on the Cora network. We specifically consider four variants of TAS-Com associated with different designs of the loss functions. Figure \ref{fig2:AblationStudy} compares all variants based on both $Q$ and $N\!M\!I$. In this figure, $L_M$ indicates that $L = L_M$, $L_R$ indicates that $L = L_R$, $Q + L_R$ indicates that $L = Q + \mu L_R$, and $L_M + N\!M\!I$ indicates that $L = L_M + \mu N\!M\!I$, where $N\!M\!I$ refers to the attributive similarity loss with respect to the human-labeled communities (without using Algorithm \ref{alg:refinement}). Finally, $L_M + L_R$ indicates the proposed loss function in Equation~\eqref{equ:L}.

Figure \ref{fig2:AblationStudy} shows that $L_M + L_R$ outperforms $Q + L_R$ and $L_M + N\!M\!I$, confirming the importance of both $L_M$ and $L_R$ for the overall loss $L$ to be effective. While $L_M$ alone yields slightly better $Q$ value than $L_M + L_R$, it results in significantly lower $N\!M\!I$. This is because $L_M$ focuses only on topological similarity. Similarly, $L_R$ achieves a slightly higher $N\!M\!I$ but a noticeably lower $Q$ score compared to $L_M + L_R$, as it considers only attributive similarity. Hence, it is crucial to balance topological and attributive similarities by including both $L_M$ and $L_R$ in the overall loss $L$. This ablation study confirms the effectiveness of the proposed loss function $L$.

\section{Conclusion}\label{Conclusion}

In this paper, we developed TAS-Com, a novel GCN-based approach powered by a newly designed loss function to effectively train GCNs to extract high-level node embeddings in attributed social networks. Our new loss function leverages the highly effective and scalable Leiden algorithm to identify community structures with globally optimal modularity ($Q$). Meanwhile, Leiden refines human-labeled communities, ensuring connectivity within each community and enabling GCNs to effectively learn node similarity at both the topology and attribute levels. Thanks to our new loss function design, the trained GCNs can detect community structures with a desirable balance between topological similarity (i.e., modularity ($Q$)) and attributive similarity (i.e., $N\!M\!I$). Experiments on multiple benchmark networks with varying sizes and complexities show that TAS-Com significantly outperforms 12 state-of-the-art approaches.


Future work could explore the potential integration of temporal dynamics or multi-view data to enhance the wide applicability of TAS-Com, including dynamic social networks. 

\bibliographystyle{named}
\bibliography{ijcai25}

\begin{thebibliography}{}

\bibitem[\protect\citeauthoryear{Bhowmick \bgroup \em et al.\egroup }{2024}]{bhowmick2024dgcluster}
Aritra Bhowmick, Mert Kosan, Zexi Huang, Ambuj Singh, and Sourav Medya.
\newblock Dgcluster: A neural framework for attributed graph clustering via modularity maximization.
\newblock In {\em Proceedings of the AAAI Conference on Artificial Intelligence}, volume~38, pages 11069--11077, 2024.

\bibitem[\protect\citeauthoryear{Bianchi \bgroup \em et al.\egroup }{2020}]{bianchi2020spectral}
Filippo~Maria Bianchi, Daniele Grattarola, and Cesare Alippi.
\newblock Spectral clustering with graph neural networks for graph pooling.
\newblock In {\em International conference on machine learning}, pages 874--883. PMLR, 2020.

\bibitem[\protect\citeauthoryear{Blondel \bgroup \em et al.\egroup }{2008}]{blondel2008fast}
Vincent~D Blondel, Jean-Loup Guillaume, Renaud Lambiotte, and Etienne Lefebvre.
\newblock Fast unfolding of communities in large networks.
\newblock {\em Journal of statistical mechanics: theory and experiment}, 2008(10), 2008.

\bibitem[\protect\citeauthoryear{Bo \bgroup \em et al.\egroup }{2020}]{bo2020structural}
Deyu Bo, Xiao Wang, Chuan Shi, Meiqi Zhu, Emiao Lu, and Peng Cui.
\newblock Structural deep clustering network.
\newblock In {\em Proceedings of the web conference 2020}, pages 1400--1410, 2020.

\bibitem[\protect\citeauthoryear{Chunaev}{2020}]{chunaev2020community}
Petr Chunaev.
\newblock Community detection in node-attributed social networks: a survey.
\newblock {\em Computer Science Review}, 37:100286, 2020.

\bibitem[\protect\citeauthoryear{Combe \bgroup \em et al.\egroup }{2015}]{combe2015louvain}
David Combe, Christine Largeron, Mathias G{\'e}ry, and El{\H{o}}d Egyed-Zsigmond.
\newblock I-louvain: An attributed graph clustering method.
\newblock In {\em Advances in Intelligent Data Analysis XIV: 14th International Symposium. Proceedings 14}, pages 181--192. Springer, 2015.

\bibitem[\protect\citeauthoryear{de Silva \bgroup \em et al.\egroup }{2022}]{de2022genetic}
Anjali de~Silva, Aaron Chen, Hui Ma, and Mohammad Nekooei.
\newblock Genetic algorithm with a novel leiden-based mutation operator for community detection.
\newblock In {\em Advances in Artificial Intelligence: 35th Australasian Joint Conference, Proceedings}, pages 252--265. Springer, 2022.

\bibitem[\protect\citeauthoryear{de Silva \bgroup \em et al.\egroup }{2023}]{de2023leiden}
Anjali de~Silva, Gang Chen, Hui Ma, and Seyed~Mohammad Nekooei.
\newblock Leiden fitness-based genetic algorithm with niching for community detection in large social networks.
\newblock In {\em Pacific Rim International Conference on Artificial Intelligence}, pages 423--435. Springer, 2023.

\bibitem[\protect\citeauthoryear{Guo \bgroup \em et al.\egroup }{2024}]{guo2024evolutionary}
Kun Guo, Zhanhong Chen, Zhiyong Yu, Kai Chen, and Wenzhong Guo.
\newblock Evolutionary computing empowered community detection in attributed networks.
\newblock {\em IEEE Communications Magazine}, 62(5):22--26, 2024.

\bibitem[\protect\citeauthoryear{He \bgroup \em et al.\egroup }{2021a}]{he2021self}
Dongxiao He, Shuai Li, Di~Jin 0001, Pengfei Jiao, and Yuxiao Huang.
\newblock Self-guided community detection on networks with missing edges.
\newblock In {\em IJCAI}, pages 3508--3514, 2021.

\bibitem[\protect\citeauthoryear{He \bgroup \em et al.\egroup }{2021b}]{he2021community}
Dongxiao He, Yue Song, Di~Jin, Zhiyong Feng, Binbin Zhang, Zhizhi Yu, and Weixiong Zhang.
\newblock Community-centric graph convolutional network for unsupervised community detection.
\newblock In {\em Proceedings of the twenty-ninth international conference on international joint conferences on artificial intelligence}, pages 3515--3521, 2021.

\bibitem[\protect\citeauthoryear{He \bgroup \em et al.\egroup }{2022}]{he2022semi}
Chaobo He, Yulong Zheng, Junwei Cheng, Yong Tang, Guohua Chen, and Hai Liu.
\newblock Semi-supervised overlapping community detection in attributed graph with graph convolutional autoencoder.
\newblock {\em Information Sciences}, 608:1464--1479, 2022.

\bibitem[\protect\citeauthoryear{Jin \bgroup \em et al.\egroup }{2019}]{jin2019graph}
Di~Jin, Ziyang Liu, Weihao Li, Dongxiao He, and Weixiong Zhang.
\newblock Graph convolutional networks meet markov random fields: Semi-supervised community detection in attribute networks.
\newblock In {\em Proceedings of the AAAI conference on artificial intelligence}, volume~33, pages 152--159, 2019.

\bibitem[\protect\citeauthoryear{Jin \bgroup \em et al.\egroup }{2021}]{jin2021survey}
Di~Jin, Zhizhi Yu, Pengfei Jiao, Shirui Pan, Dongxiao He, Jia Wu, S~Yu Philip, and Weixiong Zhang.
\newblock A survey of community detection approaches: From statistical modeling to deep learning.
\newblock {\em IEEE Transactions on Knowledge and Data Engineering}, 35(2):1149--1170, 2021.

\bibitem[\protect\citeauthoryear{Ju \bgroup \em et al.\egroup }{2023}]{ju2023glcc}
Wei Ju, Yiyang Gu, Binqi Chen, Gongbo Sun, Yifang Qin, Xingyuming Liu, Xiao Luo, and Ming Zhang.
\newblock Glcc: A general framework for graph-level clustering.
\newblock In {\em Proceedings of the AAAI conference on artificial intelligence}, volume~37, pages 4391--4399, 2023.

\bibitem[\protect\citeauthoryear{Karata{\c{s}} and {\c{S}}ahin}{2018}]{karatacs2018application}
Arzum Karata{\c{s}} and Serap {\c{S}}ahin.
\newblock Application areas of community detection: A review.
\newblock In {\em International congress on big data, deep learning and fighting cyber terrorism}, pages 65--70. IEEE, 2018.

\bibitem[\protect\citeauthoryear{Klambauer \bgroup \em et al.\egroup }{2017}]{klambauer2017self}
G{\"u}nter Klambauer, Thomas Unterthiner, Andreas Mayr, and Sepp Hochreiter.
\newblock Self-normalizing neural networks.
\newblock {\em Advances in neural information processing systems}, 30, 2017.

\bibitem[\protect\citeauthoryear{Kumar \bgroup \em et al.\egroup }{2023}]{kumar2023community}
Sanjay Kumar, Abhishek Mallik, and Sandeep~Singh Sengar.
\newblock Community detection in complex networks using stacked autoencoders and crow search algorithm.
\newblock {\em The Journal of Supercomputing}, 79(3):3329--3356, 2023.

\bibitem[\protect\citeauthoryear{Li \bgroup \em et al.\egroup }{2017}]{li2017multiobjective}
Zhangtao Li, Jing Liu, and Kai Wu.
\newblock A multiobjective evolutionary algorithm based on structural and attribute similarities for community detection in attributed networks.
\newblock {\em IEEE transactions on cybernetics}, 48(7):1963--1976, 2017.

\bibitem[\protect\citeauthoryear{Liu \bgroup \em et al.\egroup }{2022}]{liu2022deep}
Yue Liu, Wenxuan Tu, Sihang Zhou, Xinwang Liu, Linxuan Song, Xihong Yang, and En~Zhu.
\newblock Deep graph clustering via dual correlation reduction.
\newblock In {\em Proceedings of the AAAI conference on artificial intelligence}, volume~36, pages 7603--7611, 2022.

\bibitem[\protect\citeauthoryear{Liu \bgroup \em et al.\egroup }{2024}]{liu2024information}
Hongtao Liu, Jiahao Wei, Yiming Wu, and Cong Liang.
\newblock Information-enhanced deep graph clustering network.
\newblock {\em Neurocomputing}, page 127992, 2024.

\bibitem[\protect\citeauthoryear{Luo and Yan}{2020}]{luo2020adaptive}
Mengqing Luo and Hui Yan.
\newblock Adaptive attributed network embedding for community detection.
\newblock In {\em Chinese Conference on Pattern Recognition and Computer Vision}, pages 161--172. Springer, 2020.

\bibitem[\protect\citeauthoryear{Moradi \bgroup \em et al.\egroup }{2015}]{moradi2015effective}
Parham Moradi, Sajad Ahmadian, and Fardin Akhlaghian.
\newblock An effective trust-based recommendation method using a novel graph clustering algorithm.
\newblock {\em Physica A: Statistical mechanics and its applications}, 436:462--481, 2015.

\bibitem[\protect\citeauthoryear{Newman and Girvan}{2004}]{newman2004finding}
Mark~EJ Newman and Michelle Girvan.
\newblock Finding and evaluating community structure in networks.
\newblock {\em Physical review E}, 69(2):1--15, 2004.

\bibitem[\protect\citeauthoryear{Perozzi \bgroup \em et al.\egroup }{2014}]{perozzi2014deepwalk}
Bryan Perozzi, Rami Al-Rfou, and Steven Skiena.
\newblock Deepwalk: Online learning of social representations.
\newblock In {\em Proceedings of the 20th ACM SIGKDD international conference on Knowledge discovery and data mining}, pages 701--710, 2014.

\bibitem[\protect\citeauthoryear{Pizzuti and Socievole}{2018}]{pizzuti2018genetic}
Clara Pizzuti and Annalisa Socievole.
\newblock A genetic algorithm for community detection in attributed graphs.
\newblock In {\em Applications of Evolutionary Computation: 21st International Conference, Proceedings 21}, pages 159--170. Springer, 2018.

\bibitem[\protect\citeauthoryear{Sen \bgroup \em et al.\egroup }{2008}]{sen2008collective}
Prithviraj Sen, Galileo Namata, Mustafa Bilgic, Lise Getoor, Brian Galligher, and Tina Eliassi-Rad.
\newblock Collective classification in network data.
\newblock {\em AI magazine}, 29(3):93--93, 2008.

\bibitem[\protect\citeauthoryear{Shchur and G{\"u}nnemann}{2019}]{shchur2019overlapping}
Oleksandr Shchur and Stephan G{\"u}nnemann.
\newblock Overlapping community detection with graph neural networks.
\newblock {\em arXiv preprint arXiv:1909.12201}, 2019.

\bibitem[\protect\citeauthoryear{Shchur \bgroup \em et al.\egroup }{2018}]{shchur2018pitfalls}
Oleksandr Shchur, Maximilian Mumme, Aleksandar Bojchevski, and Stephan G{\"u}nnemann.
\newblock Pitfalls of graph neural network evaluation. arxiv 2018.
\newblock {\em arXiv preprint arXiv:1811.05868}, 2018.

\bibitem[\protect\citeauthoryear{Su \bgroup \em et al.\egroup }{2022}]{su2022comprehensive}
Xing Su, Shan Xue, Fanzhen Liu, Jia Wu, Jian Yang, Chuan Zhou, Wenbin Hu, Cecile Paris, Surya Nepal, Di~Jin, et~al.
\newblock A comprehensive survey on community detection with deep learning.
\newblock {\em IEEE Transactions on Neural Networks and Learning Systems}, pages 1--21, 2022.

\bibitem[\protect\citeauthoryear{Sun \bgroup \em et al.\egroup }{2019}]{sun2019vgraph}
Fan-Yun Sun, Meng Qu, Jordan Hoffmann, Chin-Wei Huang, and Jian Tang.
\newblock vgraph: A generative model for joint community detection and node representation learning.
\newblock {\em Advances in Neural Information Processing Systems}, 32, 2019.

\bibitem[\protect\citeauthoryear{Sun \bgroup \em et al.\egroup }{2020}]{sun2020network}
Heli Sun, Fang He, Jianbin Huang, Yizhou Sun, Yang Li, Chenyu Wang, Liang He, Zhongbin Sun, and Xiaolin Jia.
\newblock Network embedding for community detection in attributed networks.
\newblock {\em ACM Transactions on Knowledge Discovery from Data}, 14(3):1--25, 2020.

\bibitem[\protect\citeauthoryear{Traag \bgroup \em et al.\egroup }{2019}]{traag2019louvain}
Vincent~A Traag, Ludo Waltman, and Nees~Jan Van~Eck.
\newblock From louvain to leiden: guaranteeing well-connected communities.
\newblock {\em Scientific reports}, 9(1):1--12, 2019.

\bibitem[\protect\citeauthoryear{Tsitsulin \bgroup \em et al.\egroup }{2023}]{tsitsulin2023graph}
Anton Tsitsulin, John Palowitch, Bryan Perozzi, and Emmanuel M{\"u}ller.
\newblock Graph clustering with graph neural networks.
\newblock {\em Journal of Machine Learning Research}, 24(127):1--21, 2023.

\bibitem[\protect\citeauthoryear{Veli{\v{c}}kovi{\'c} \bgroup \em et al.\egroup }{2018}]{velivckovic2018deep}
Petar Veli{\v{c}}kovi{\'c}, William Fedus, William~L Hamilton, Pietro Li{\`o}, Yoshua Bengio, and R~Devon Hjelm.
\newblock Deep graph infomax.
\newblock {\em arXiv preprint arXiv:1809.10341}, 2018.

\bibitem[\protect\citeauthoryear{Wang \bgroup \em et al.\egroup }{2019}]{wang2019attributed}
Chun Wang, Shirui Pan, Ruiqi Hu, Guodong Long, Jing Jiang, and Chengqi Zhang.
\newblock Attributed graph clustering: A deep attentional embedding approach.
\newblock {\em arXiv preprint arXiv:1906.06532}, 2019.

\bibitem[\protect\citeauthoryear{Wang \bgroup \em et al.\egroup }{2021}]{wang2021unsupervised}
Xiaofeng Wang, Jianhua Li, Li~Yang, and Hongmei Mi.
\newblock Unsupervised learning for community detection in attributed networks based on graph convolutional network.
\newblock {\em Neurocomputing}, 456:147--155, 2021.

\bibitem[\protect\citeauthoryear{Wu \bgroup \em et al.\egroup }{2020}]{wu2020deep}
Ling Wu, Qishan Zhang, Chi-Hua Chen, Kun Guo, and Deqin Wang.
\newblock Deep learning techniques for community detection in social networks.
\newblock {\em IEEE Access}, 8:96016--96026, 2020.

\bibitem[\protect\citeauthoryear{Xia \bgroup \em et al.\egroup }{2021}]{xia2021self}
Wei Xia, Quanxue Gao, Ming Yang, and Xinbo Gao.
\newblock Self-supervised contrastive attributed graph clustering.
\newblock {\em arXiv preprint arXiv:2110.08264}, 2021.

\bibitem[\protect\citeauthoryear{Xie \bgroup \em et al.\egroup }{2021}]{xie2021effective}
Xiaoqin Xie, Mingjie Song, Chiming Liu, Jiaming Zhang, and Jiahui Li.
\newblock Effective influential community search on attributed graph.
\newblock {\em Neurocomputing}, 444:111--125, 2021.

\bibitem[\protect\citeauthoryear{Yang \bgroup \em et al.\egroup }{2016}]{yang2016modularity}
Liang Yang, Xiaochun Cao, Dongxiao He, Chuan Wang, Xiao Wang, and Weixiong Zhang.
\newblock Modularity based community detection with deep learning.
\newblock In {\em IJCAI}, volume~16, pages 2252--2258, 2016.

\bibitem[\protect\citeauthoryear{Yang \bgroup \em et al.\egroup }{2023}]{yang2023cluster}
Xihong Yang, Yue Liu, Sihang Zhou, Siwei Wang, Wenxuan Tu, Qun Zheng, Xinwang Liu, Liming Fang, and En~Zhu.
\newblock Cluster-guided contrastive graph clustering network.
\newblock In {\em Proceedings of the AAAI conference on artificial intelligence}, volume~37, pages 10834--10842, 2023.

\bibitem[\protect\citeauthoryear{Ying \bgroup \em et al.\egroup }{2018}]{ying2018graph}
Rex Ying, Ruining He, Kaifeng Chen, Pong Eksombatchai, William~L Hamilton, and Jure Leskovec.
\newblock Graph convolutional neural networks for web-scale recommender systems.
\newblock In {\em Proceedings of the 24th ACM SIGKDD international conference on knowledge discovery \& data mining}, pages 974--983, 2018.

\bibitem[\protect\citeauthoryear{Zhang \bgroup \em et al.\egroup }{1996}]{zhang1996birch}
Tian Zhang, Raghu Ramakrishnan, and Miron Livny.
\newblock Birch: an efficient data clustering method for very large databases.
\newblock {\em ACM sigmod record}, 25(2):103--114, 1996.

\bibitem[\protect\citeauthoryear{Zhou \bgroup \em et al.\egroup }{2023}]{zhou2023community}
Xinchuang Zhou, Lingtao Su, Xiangju Li, Zhongying Zhao, and Chao Li.
\newblock Community detection based on unsupervised attributed network embedding.
\newblock {\em Expert Systems with Applications}, 213:118937, 2023.

\bibitem[\protect\citeauthoryear{Zhu \bgroup \em et al.\egroup }{2024}]{zhu2024every}
Pengfei Zhu, Qian Wang, Yu~Wang, Jialu Li, and Qinghua Hu.
\newblock Every node is different: Dynamically fusing self-supervised tasks for attributed graph clustering.
\newblock In {\em Proceedings of the AAAI Conference on Artificial Intelligence}, volume~38, pages 17184--17192, 2024.

\end{thebibliography}

\clearpage

\appendix
\section*{Appendices}
\section{GCN training process in TAS-Com}

Algorithm \ref{alg:TAS-Com} presents the overall design of the proposed TAS-Com approach. The two key steps in TAS-Com are obtaining $CS_L$ (lines 2--5) and $CS_R$ (line 8) in Algorithm \ref{alg:TAS-Com}. They are used to generate the two loss functions: $L_M$ and $L_R$, which are combined to build the overall loss function $L = L_M + \mu L_R$. Subsequently, the Graph Convolutional Neural Network (GCN) model is trained under the guidance of $L$ through the stochastic gradient descent method. Specifically, the GCN model is trained for 300 epochs through backpropagation. The Adam optimizer is used to update all trainable parameters of the GCN model in the direction of minimizing $L$. The training process of the GCN model corresponds to lines 10--15 in Algorithm \ref{alg:TAS-Com}. The trained GCN model is used to get the node embedding ($X^{(e)}$) for a given network $N$ (line 16 in Algorithm \ref{alg:TAS-Com}). Afterward, the final community structure $CS$ is obtained by applying the BIRCH clustering algorithm on $X^{(e)}$ (line 17 in Algorithm \ref{alg:TAS-Com}). 

\setcounter{algorithm}{1}
\begin{algorithm}[h]
\caption{The TAS-Com approach for community detection in attributed networks}
\label{alg:TAS-Com}
\textbf{Input}: Social Network $N$; Human-labeled communities $CS_O$; Hyperparameters $\mu, threshold$; Number of epochs $N_e$\\
\textbf{Output}: Optimal community structure $CS$
\begin{algorithmic}[1] 
\STATE \textbf{// Obtain $CS_L$} \COMMENT{Refer Subsection Modularity-based similarity loss ($L_M$)}
\FOR{ $i = 1 $ to $n$}
    \STATE $CS_l \longleftarrow Leiden(N)$
    \STATE $N\!M\!I(CS_l) = N\!M\!I(CS_l, CS_O)$
\ENDFOR
\STATE $CS_L \longleftarrow $ $CS_l$ having the highest $N\!M\!I(CS_l)$
\STATE \textbf{// Obtain $CS_R$} \COMMENT{Refer Subsection Refined human-label-based similarity loss ($L_R$)}
\STATE $CS_R \longleftarrow $ Refine $CS_O$ \COMMENT{Refer Algorithm 1}
\STATE \textbf{// Train GCN model}
\FOR{ $epoch = 1$ to $N_e$ }
    \STATE $X^{(e)} \longleftarrow $ Obtain $X^{(e)}$ from GCN model
    \STATE Calculate the loss $L = L_M + \mu L_R$ according to Equations (7) and (9)
    \STATE Compute the gradients of $L$ with respect to the model parameters
    \STATE Update the model's weights using the Adam optimizer
\ENDFOR
\STATE $X^{(e)} \longleftarrow $ Obtain the final node embedding via the trained GCN model
\STATE $CS \longleftarrow $ Obtain optimal community structure $CS$ by applying $BIRCH(X^{(e)})$ 

\STATE \textbf{return} $CS$
\end{algorithmic}
\end{algorithm}

\section{Parameter sensitivity analysis}
 This appendix performs the sensitivity analysis of the hyperparameters of the proposed TAS-Com approach. Two hyperparameters will be studied, including (1) $\mu$ that controls the influence of refined human-labeled-based similarity loss ($L_R$) in overall loss $L$ and (2) $threshold$ used in Algorithm 1, corresponding to the second step of refining the human-labeled communities.  

\subsection{Sensitivity analysis on $\mu$}

\begin{figure}[h]
    \centering
    \includegraphics[width=1.0\linewidth]{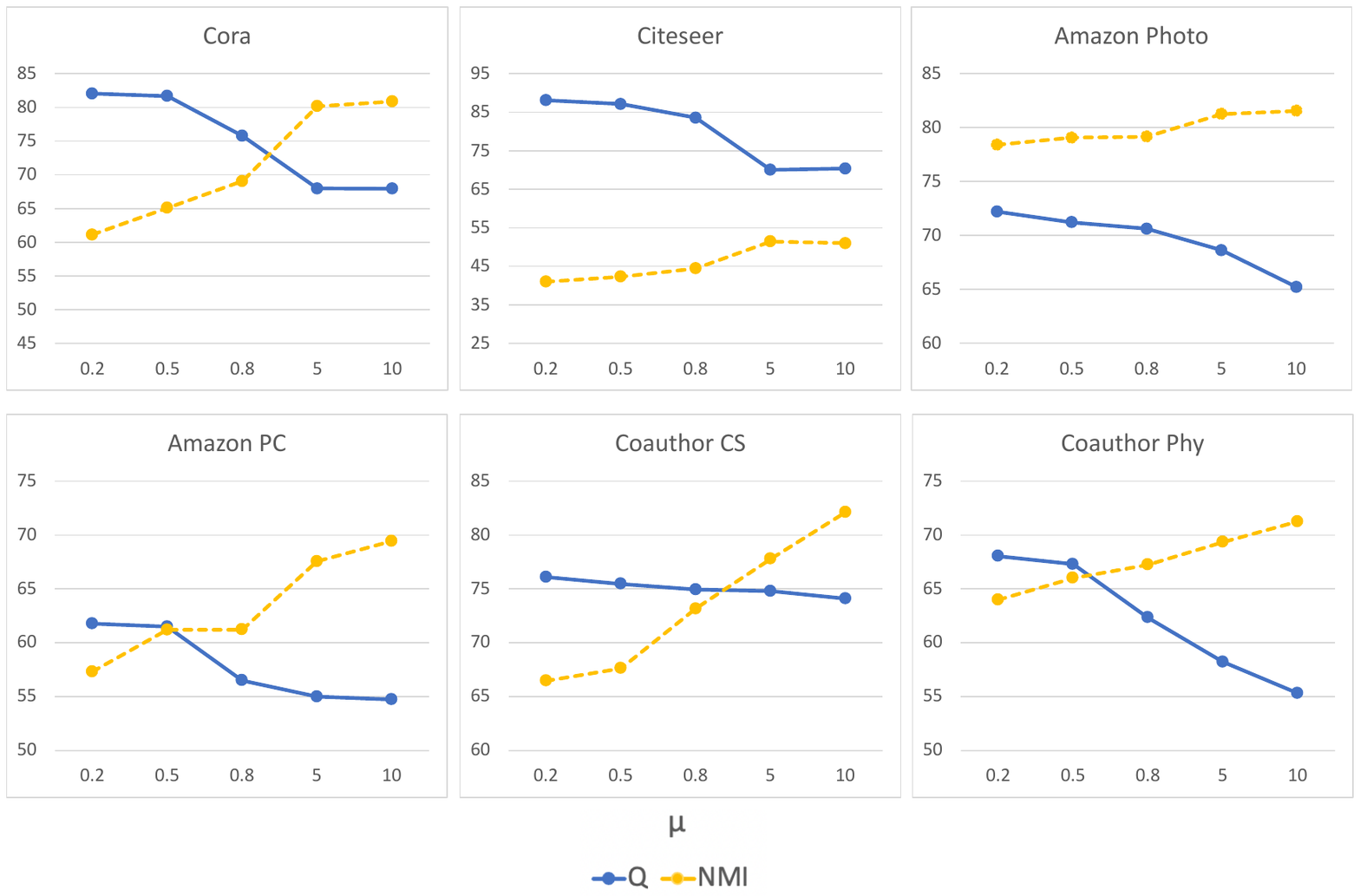}
    \caption{Sensitivity analysis of parameter $\mu$. Modularity $Q$ and $N\!M\!I$ scores are scaled by a factor of 100 for ease of reading.}
    \label{fig:muSensitivity}
\end{figure}

In the overall loss function $L = L_M + \mu L_R$, $\mu$ plays a crucial role as it controls the influence of attributive similarity among nodes while performing community detection (CD). We conduct the experiments with 5 different values for $\mu >0$, including $\{0.2, 0.5, 0.8, 5, 10\}$. Fig. \ref{fig:muSensitivity} demonstrates how the performance of TAS-Com varies with different settings of $\mu$ on each benchmark network in terms of both modularity $Q$ and $N\!M\!I$.

Each graph in Fig. \ref{fig:muSensitivity} contains two plots showing the values of $Q$ and $N\!M\!I$ across five different settings of $\mu$. According to the plotted graphs, we can see that increasing $\mu$ leads directly to increased $N\!M\!I$. This implies that increasing the influence of attributive similarity in the overall loss function helps to increase the quality of the community structure in terms of $N\!M\!I$. Meanwhile, as $\mu$ increases, $Q$ continues to deteriorate. It is also worthwhile to note that the suitable settings of $\mu$ depend on the network $N$. In practice, TAS-Com can be used multiple times, preferably in parallel and under different settings of $\mu$, to process the same network. Consequently, a group of different community structures can be created for the end users to choose, based on their individual preferences. Furthermore, it is worth noting that even in the absence of ground-truth labels, practitioners can tune TAS-Com parameters (e.g., $\mu$) using internal quality metrics such as modularity $Q$ or conductance, without requiring any labeled data. These parameters can be iteratively adjusted to maximize the chosen metrics, enabling effective unsupervised optimization.

For the experiment results reported in the section titled ``Results and Discussion", the specific settings of $\mu$ adopted by TAS-Com with respect to each benchmark network are summarized in Table \ref{tab:musettings}.

\begin{table}[H]
   \centering
   \begin{tabular}{cc}
   \toprule
   \textbf{Network} & $\mu$ \\ 
   \midrule
    Cora & 0.5  \\
    Citeseer & 0.2 \\
    Amazon Photo & 0.2 \\
    Amazon PC & 0.5  \\
    Coauthor CS & 10 \\
    Coauthor Phy & 0.5 \\
   \bottomrule
   \end{tabular}
   \caption{Settings of $\mu$ for TAS-Com across all experimented benchmark networks.}\label{tab:musettings}
\end{table}

\subsection{Sensitivity analysis on $threshold$}

Step 2 in Algorithm 1 (i.e., the step to merge sub-communities) relies on $threshold$ to determine when the merging process should be terminated. We considered two different ways to configure $threshold$: (1) $threshold=\frac{|CC(N_L)|}{2}$ and (2) $threshold=|CC(N_L)|$, where $CC(N_L) = \{C_1, ...., C_k\}$ refers to the set of connected components of the sub-network $N_L$ that contains all nodes of network $N$ in $CS_L$. According to our experiment results, we observe that the second way significantly reduces $N\!M\!I$ without noticeably improving $Q$ on most of the experimented benchmark networks. Table \ref{tab:threshold settings} specifically shows the performance differences between the above two ways of setting $threshold$ on two benchmark networks, i.e., Cora and Citeseer. According to the presented results, for both networks, $N\!M\!I$ has significantly dropped with only a slight increment in $Q$ upon using the second way. Further, these results imply that the first condition, i.e., $threshold=\frac{|CC(N_L)|}{2}$, effectively reduces disconnected communities while preserving alignment with human-labeled community structures. Driven by the aim to balance $Q$ and $N\!M\!I$, we decide to adopt the first way and set $threshold=\frac{|CC(N_L)|}{2}$.

\begin{table}[H]
   \centering
   \setlength{\tabcolsep}{2pt}
   \begin{tabular}{c|ccc|ccc}
   \toprule
   Network & \multicolumn{3}{|c|}{$threshold=\frac{|CC(N_L)|}{2}$} & \multicolumn{3}{|c}{$threshold=|CC(N_L)|$} \\ 
    &  & $Q$ & $N\!M\!I$ &  & $Q$ & $N\!M\!I$ \\
    \midrule
    Cora &  & 81.7 & 65.1 &  & 81.9 & 48.7 \\
    Citeseer &  & 88.1 & 41.0 &  & 88.24 & 36.76 \\
   \bottomrule
   \end{tabular}
   \caption{Performance comparison on Cora and Citeseer networks for two $threshold$ setting methods.}\label{tab:threshold settings}
\end{table}

\section{Transformation of node embedding}

The node embedding $X^{(3)}$ obtained from the third convolutional layer (i.e., the last convolutional layer) of our GCN network introduced in the section titled ``Proposed Method" may potentially contain negative values. It prevents us from training the GCN network to minimize the loss function $L$ where the loss targets are matrices with non-negative entries. To address this issue, following DGCluster \cite{bhowmick2024dgcluster}, a series of transformation steps will be performed on $X^{(3)}$ to ensure that the final node embedding $X^{(e)}$ only has non-negative entries. Specifically, the first transformation step normalizes every entry of $X^{(3)}$ according to Equation \eqref{equ:firstnorm} below:

\begin{equation}\label{equ:firstnorm}
    \overline{X}_{ij}^{(3)} = \frac{X_{ij}^{(3)}}{X_{i.}}.
\end{equation}

The above equation helps prevent the vanishing gradient issue due to the subsequent use of the $\tanh(\cdot)$ activation function as shown in Equation \eqref{equ:tanh} below: 

\begin{equation}\label{equ:tanh}
    \hat{X}_{ij}^{(3)} = \tanh(\overline{X}_{ij}^{(3)}).
\end{equation}

Afterwards, Equation \eqref{equ:L2Normalization} is performed to generate the final node embedding $X^{(e)}$ that will always fall into the positive coordinate space. It also simplifies the computation of the cosine similarity between every pair of nodes in the network \cite{bhowmick2024dgcluster}. Particularly, let $X_i$ and $X_j$ represent the $i$-th and $j$-th rows of $X^{(e)}$. In other words, $X_i$ and $X_j$ are the embeddings of node $i$ and node $j$, respectively. Then, the cosine similarity between the two nodes can be computed straightforwardly as $\cos(i,j) = X_u^T \cdot X_v$, which is used directly by our proposed loss function $L$.

\begin{equation}\label{equ:L2Normalization}
    X_{ij}^{(e)} = \frac{(\hat{X}_{ij}^{(3)})^2}{\sqrt{\sum_j (\hat{X}_{ij}^{(3)})^2}}.
\end{equation}


\section{Modularity ($Q$) of human-labeled communities}

As explained in the introduction section, human-labeled communities were generated based on solely the node attribute information. Hence, they often contain disconnected nodes, hurting the quality of the community structure in terms of modularity $Q$. For example, concerning the Cora network, the quality of the human-labeled communities in terms of $Q$ is $64.01$. When each isolated sub-network in the human-labeled communities is treated as a separate community, the quality of the new community structure in terms of $Q$ is increased to $67.43$. This result suggests $Q$ can be substantially improved by avoiding isolated sub-networks in human-labeled communities. In fact, TAS-Com achieved a much higher $Q$ score of $81.7$ on Cora, confirming the necessity of explicitly considering the topological structures as a crucial factor for identifying high-quality community structures. 

\section{Importance of merging sub-communities in Algorithm 1}

This appendix compares three different community structures: (1) human-labeled communities $CS_O$; (2) community structure $CS_L$ obtained after performing step 1 in Algorithm 1; and (3) community structure $CS_R$ obtained after performing step 2 in Algorithm 1. In particular, $CS_L$ is obtained by applying Leiden \cite{traag2019louvain} to create sub-communities within each human-labelled community in $CS_O$. $CS_R$ is obtained by merging sub-communities in $CS_L$.

\begin{table*}[htb!]
   \centering
   \setlength{\tabcolsep}{3pt}
   \begin{tabular}{c|ccc|ccc|ccc}
   \toprule
   \textbf{Network} & \multicolumn{3}{c|}{Human-labeled Communities} & \multicolumn{3}{c|}{Step 1 in Algorithm 1} & \multicolumn{3}{c}{Step 2 in Algorithm 1} \\
    & $Q$ & $N\!M\!I$ & $|CS_O|$ & $Q$ & $N\!M\!I$ & $|CS_L|$ & $Q$ & $N\!M\!I$ & $|CS_R|$ \\
    \midrule
    Cora & 64.0 & 1.0 & 7 & 67.0 & 52.8 & 385 & 67.4 & 75.5 & 146 \\
    Citeseer & 53.9 & 1.0 & 6 & 65.6 & 43.9 & 1136 & 65.4 & 50.5 & 539 \\
   \bottomrule
   \end{tabular}
   \caption{Performance comparison that demonstrates the necessity of merging sub-communities in step 2 of Algorithm 1.}\label{tab:Leidenvsgroundtruth}
\end{table*} 

Table \ref{tab:Leidenvsgroundtruth} compares the three community structures in terms of $Q$, $N\!M\!I$, and the number of communities with respect to the Cora and Citeseer networks. Our results show that the Leiden algorithm frequently produces numerous small sub-communities ($|CS_L|$) that significantly deviate from human-labeled communities ($|CS_O|$) because Leiden relies solely on topology, while human-labeled communities consider node attributes. Due to this reason, $CS_L$ exhibits poor $N\!M\!I$. In comparison, $CS_R$ substantially improves $N\!M\!I$ while reducing the number of communities ($|CS_R|$). Meanwhile, on both networks, $CS_R$ achieved similar $Q$ scores as $CS_L$. This observation highlights the importance of merging closely connected sub-communities, leading to community structures with high modularity while maintaining high $N\!M\!I$.


\section{Analysis of community structures generated by TAS-Com}

As explained in the subsection titled ``Further analysis", TAS-Com was shown to generate less isolated sub-networks compared to DGCluster. In this appendix, we further investigate the best community structure identified by TAS-Com to determine whether the presence of these isolated sub-networks hurts the effectiveness of the solution. For this purpose, we create a modified approach called \emph{Modified-TAS-Com} that turns every isolated sub-network within any community identified by TAS-Com into separate communities. Table \ref{tab:optimalSolutionAnalysis} presents the performance comparison between TAS-Com and Modified-TAS-Com in terms of $Q$ and $N\!M\!I$. According to the reported results, Modified-TAS-Com slightly improves the modularity $Q$ while noticeably reducing $N\!M\!I$. Hence, we can confidently conclude that, while TAS-Com may generate a small number of isolated sub-networks, it is not necessary to treat these isolated sub-networks as separate communities. In fact, compared to Modified-TAS-Com, TAS-Com can achieve a much better trade-off between $Q$ and $N\!M\!I$, outperforming all competing approaches on most of the benchmark networks. Hence, we propose TAS-Com in the main paper.

\begin{table}[h]
   \centering
   \begin{tabular}{ccccc}
   \toprule
   \textbf{Network} & \multicolumn{2}{c}{TAS-Com} & \multicolumn{2}{c}{Modified-TAS-Com} \\ 
   \midrule
    & $Q$ & $N\!M\!I$ & $Q$ & $N\!M\!I$ \\
    \midrule
    Cora & 81.7 & 65.1 & 81.8 & 63.4 \\
    Citeseer & 88.1 & 41.0 & 88.3 & 38.3 \\
    Amazon Photo & 72.2 & 78.4 & 72.5 & 73.4 \\
    Amazon PC & 61.5 & 61.2 & 61.5 & 59.3 \\
    Coauthor CS & 74.1 & 82.1 & 74.4 & 79.2 \\
    Coauthor Phy & 67.3 & 66.0 & 67.9 & 62.5 \\
   \bottomrule
   \end{tabular}
   \caption{Performance comparison between the TAS-Com and Modified-TAS-Com in terms of $Q$ and $N\!M\!I$. Modified-TAS-Com turns every isolated sub-network within the same community identified by TAS-Com as separate communities.}\label{tab:optimalSolutionAnalysis}
\end{table}

\section{Description of baseline approaches}

As explained in the subsection titled ``Baseline approaches", the performance of TAS-Com is compared with 11 state-of-the-art approaches. In this appendix, we present further details about these baseline approaches. Since the influence of topological similarity and attributive similarity presents the main concern in developing the new TAS-Com approach, we consider baseline approaches in three main categories as specified in the next three subsections.

\subsection{Approaches using either attribute or topological information alone for community detection}

k-m(feat) (i.e., k-means based only on features) is an approach that only considers the attribute information while identifying the communities in a given network $N$. This approach uses the local Lloyd algorithm with the k-means++ seeding strategy \cite{tsitsulin2023graph} to perform CD. DMoN \cite{tsitsulin2023graph} is a recently proposed approach that directly optimizes community assignments using modularity as its loss function. Hence, the topological information is only considered by DMoN when performing CD. It overlooks the attribute similarity while optimizing the community assignments. To demonstrate the importance of considering both topological and attributive information in detecting hidden community structures, we use these two algorithms as our competing approaches. 

\subsection{Approaches considering node similarities at both topological and attribute levels
}

k-m(DW) \cite{perozzi2014deepwalk} proposed a novel strategy of concatenating node attributes to learned node embedding in social networks. k-means(DGI) \cite{velivckovic2018deep} combines the learning of communities and node representations. DAEGC \cite{wang2019attributed} is a graph reconstruction-based community detection approach. Similarly, SDCN \cite{bo2020structural} is also a graph reconstruction-based approach that initializes the representations through k-means over auto-encoded attributes. NOCD \cite{shchur2019overlapping} is an approach that directly optimizes the negative log-likelihood of the graph reconstruction of the Bernoulli-Poisson model. 

Besides the above, DyFSS \cite{zhu2024every} proposed a novel framework for attributed graph clustering that addresses the limitations of existing self-supervised learning (SSL) approaches. DyFSS dynamically learns node-specific weights for fusing diverse SSL tasks. this node-wise adaptability is achieved through a dynamic fusion network, leveraging a Mixture of Experts (MoE) framework and a gating mechanism. Additionally, DyFSS incorporates a dual-level self-supervised strategy that combines pseudo-label and network topology information to enhance clustering performance. 

Furthermore, DGCluster is widely considered as one of the most effective approaches in this category \cite{bhowmick2024dgcluster}, which was recently proposed for CD in attributed networks. Since these approaches jointly consider the topological and attributive information in the learning process, we choose these approaches as our competing approaches.

\subsection{Approaches utilizing graph pooling techniques}

Graph pooling techniques are designed to capture hierarchical structures by reducing the size of a graph while preserving its important features, such as topological and attributive information. DiffPool \cite{ying2018graph} is a primary graph pooling approach, while MinCutPool \cite{bianchi2020spectral} is a deep pooling approach that orthogonalizes the community representations. Additionally, Ortho \cite{bianchi2020spectral} is proposed as another variant of MinCutPool that conducts community orthogonalization without following any graph-related objectives.  During the pooling process, these techniques mainly focus on the graph's topological integrity. By comparing these approaches with one that leverages both topology and node attributes, we can effectively assess how well each method preserves critical structural information. It motivates us to select these approaches as our comparison baselines.

\section{Statistical comparison of TAS-Com and DGCluster performance}

Statistical tests have been conducted to verify whether TAS-Com can outperform DGCluster with statistical significance. In particular, we conduct the Wilcoxon rank sum test at a significance level of 0.05 to compare the performance between TAS-Com and DGCluster. As a key hyperparamter of DGCluster, $\lambda$ is recommended to follow two specific settings, i.e., $\lambda = \{0.2, 0.8\}$, in \cite{bhowmick2024dgcluster}. Accordingly, TAS-Com is statistically compared with DGCluster under both settings of $\lambda$. The corresponding results are reported in Tables \ref{tab:statAnalysis_Q} and \ref{tab:statAnalysis_NMI}. In both tables, we use symbols $+/\mkern-6mu=$ to indicate that TAS-Com either performed significantly better than or similarly as DGCluster($\lambda=0.2$) and DGCluster($\lambda=0.8$) in terms of $Q$ and $N\!M\!I$ respectively. As evidenced by the results reported in Tables \ref{tab:statAnalysis_Q} and \ref{tab:statAnalysis_NMI}, TAS-Com significantly outperforms DGCluster. Given that DGCluster outperforms other competing approaches studied in this paper, we can confidently conclude that TAS-Com surpasses all competing methods. 

\begin{table}[H]
   \scriptsize
   \centering
   \begin{tabular}{cccc}
   \toprule
   \textbf{Network} & \multicolumn{2}{c}{\textbf{DGCluster} \cite{bhowmick2024dgcluster}} & \textbf{TAS-Com} \\ 
    & $\lambda = 0.2$ & $\lambda = 0.8$ & \\ 
   \midrule
    Cora & 80.8 & 78.6 & \textbf{81.7} (+)(+) \\
    Citeseer & 87.4 & 86.3 & \textbf{88.1} (+)(+) \\
    Amazon Photo & 71.6 & 71.6 & \textbf{72.2} (+)(+) \\
    Amazon PC & \textbf{61.5} & 60.3 & \textbf{61.5} (=)(+) \\
    Coauthor CS & \textbf{74.2} & 73.3 & 74.1 (=)(+) \\
    Coauthor Phy & \textbf{67.3} & 66.0 & \textbf{67.3} (=)(+)\\
   \bottomrule
   \end{tabular}
   \caption{Statistical comparison of TAS-Com and DGCluster in terms of modularity $Q$.}\label{tab:statAnalysis_Q}
\end{table} 

\begin{table}[H]
   \scriptsize
   \centering
   \begin{tabular}{cccc}
   \toprule
   \textbf{Network} & \multicolumn{2}{c}{\textbf{DGCluster} \cite{bhowmick2024dgcluster}} & \textbf{TAS-Com} \\ 
    & $\lambda = 0.2$ & $\lambda = 0.8$ & \\ 
   \midrule
    Cora & 53.0 & 62.1 & \textbf{65.1} (+)(+) \\
    Citeseer & 30.3  & \textbf{41.0}  & \textbf{41.0} (+)(=) \\
    Amazon Photo & 73.0 & 77.3 & \textbf{78.4} (+)(+) \\
    Amazon PC & 53.8 & 60.4 & \textbf{61.2} (+)(+) \\
    Coauthor CS & 76.1 & \textbf{82.1} & \textbf{82.1} (+)(=) \\
    Coauthor Phy & 59.0 & 65.7 & \textbf{66.0} (+)(+)\\
   \bottomrule
   \end{tabular}
   \caption{Statistical comparison of TAS-Com and DGCluster in terms of $N\!M\!I$.}\label{tab:statAnalysis_NMI}
\end{table} 

\section{Performance evaluation using additional metrics}

To further assess TAS-Com's performance against state-of-the-art algorithms, we use two additional metrics: Conductance ($Con$), a topology-based measure, and F1 score ($F1$), an attribute similarity-based measure \cite{bhowmick2024dgcluster}. Table \ref{tab:performanceonmoremetrics} reports the average $Con$ and $F1$ values across 10 independent runs for each algorithm on all benchmark networks. A lower $Con$ value (denoted by $\downarrow$) indicates a better-defined community structure with fewer external connections relative to internal ones, while a higher $F1$ (denoted by $\uparrow$) reflects greater accuracy in identifying community memberships, signifying a superior community structure.

As shown in Table \ref{tab:performanceonmoremetrics}, TAS-Com outperforms all state-of-the-art algorithms on the Cora, Amazon Photo, and Amazon PC networks in both $Con$ and $F1$ metrics. Although TAS-Com shows slightly higher $Con$ values for the Coauthor Phy network than the DGCluster, it achieves significantly better $F1$ scores. While TAS-Com performs marginally worse than DGCluster on Coauthor CS in both metrics, it still outperforms all other competing algorithms on this network. Overall, TAS-Com demonstrates substantial performance improvements across multiple benchmark networks in terms of $Con$ and $F1$, reaffirming its effectiveness as a highly competitive approach for CD in attributed networks.    

\begin{table*}[htb!]
   \centering
   \resizebox{\textwidth}{!}{
   \begin{tabular}{ccccccccccccc}
   \toprule
   Approach & \multicolumn{2}{c}{Cora} & \multicolumn{2}{c}{Citeseer} & \multicolumn{2}{c}{Amazon Photo} & \multicolumn{2}{c}{Amazon PC} & \multicolumn{2}{c}{Coauthor CS} & \multicolumn{2}{c}{Coauthor Phy}\\ 
   \midrule
    & $\downarrow Con$ & $\uparrow F1$ & $\downarrow Con$ & $\uparrow F1$ & $\downarrow Con$ & $\uparrow F1$ & $\downarrow Con$ & $\uparrow F1$ & $\downarrow Con$ & $\uparrow F1$ & $\downarrow Con$ & $\uparrow F1$ \\
    \midrule
    k-m(feat) & 61.7 & 27.0 & 60.5 & 29.2 & 79.6 & 19.5 & 84.5 & 19.2 & 49.1 & 39.4 & 57.0 & 42.9 \\ 
    \midrule
    k-m(DW) & 62.1 & 24.8 & 68.1 & 24.8 & 60.6 & 33.8 & 67.6 & 22.7 & 33.1 & 61.2 & 44.7 & 24.3 \\
   \midrule
    SDCN & 37.5 & 29.9 & 20.0 & 41.9 & 19.7 & 45.1 & 25.1 & 45.2 & 33.0 & 54.7 & 32.1 & 39.9 \\
   \midrule
   DAEGC & 56.8 & 13.6 & 47.6 & 18.0 & 19.3 & 45.0 & 39.0 & 37.3 & 39.4 & 32.4 & N/A & N/A \\
   \midrule
   k-m(DGI) & 28.0 & 40.1 & 17.5 & 39.4 & 51.5 & 23.6 & 61.9 & 15.0 & 35.1 & 51.9 & 38.6 & 30.6 \\
   \midrule
   NOCD & 14.7 & 36.7 & 6.8 & 24.1 & 13.7 & 60.2 & 26.4 & 37.8 & 20.9 & 56.4 & 25.7 & 28.7 \\
   \midrule
   DiffPool & 26.1 & 34.4 & 26.0 & 23.5 & 26.5 & 41.8 & 35.6 & 38.3 & 33.6 & 34.4 & N/A & N/A \\
   \midrule
   MinCutPool & 23.3 & 25.0 & 14.1 & 20.1 & N/C & N/C & N/C & N/C & 22.7 & 47.8 & 27.8 & 24.9 \\
   \midrule
   Ortho & 28.0 & 26.6 & 18.4 & 20.5 & N/C & N/C & N/C & N/C & 27.8 & 46.1 & 33.0 & 23.7 \\
   \midrule
   DMoN & 12.2 & 48.8 & 5.1 & 43.2 & 12.7 & 61.0 & 18.0 & 45.4 & 17.5 & 59.8 & \textbf{18.8} & 42.4 \\
   \midrule
    DyFSS & 10.0 & 68.1 & 4.4 & \textbf{64.7} & 19.0 & 50.8 & 45.9 & 26.2 & 25.45 & 72.7 & 21.4 & 48.6 \\
    \midrule
   DGCluster($\lambda = 0.2$) & 9.7 & 43.5 & 4.1 & 22.2 & 8.6 & 70.7 & \textbf{17.7} & 49.5 & \textbf{15.3} & 77.3 & 22.3 & 41.9 \\
   DGCluster($\lambda = 0.8$) & 14.5 & 54.5 & 6.5 & 32.2 & 12.4 & 75.9 & 27.3 & 52.2 & 18.1 & \textbf{83.5} & 21.3 & 49.2 \\
   \midrule
   \textbf{TAS-Com} & \textbf{8.8} & \textbf{56.9} & \textbf{3.8} & \underline{32.3} & \textbf{8.1} & \textbf{76.7} & \textbf{17.7} & \textbf{54.1} & \underline{15.6} & \underline{83.2} & \underline{21.2} & \textbf{50.9}\\
   \bottomrule
   \end{tabular}}
   \caption{Performance comparison in terms of conductance ($Con$) and F1 score ($F1$) (results are multiplied by 100) between TAS-Com and state-of-the-art approaches. The best and second-best results of $Con$ and $F1$ are bolded and underlined, respectively. N/A: Not Available, and N/C: Not Converged.} \label{tab:performanceonmoremetrics}
\end{table*}

\section{Performance comparison between TAS-Com and more state-of-the-art algorithms}

Table \ref{tab:performance} presents the performance outcomes of additional competing algorithms developed using graph pooling techniques. The results demonstrate that TAS-Com consistently achieves superior performance compared to both MinCutPool and Ortho algorithms with respect to $Q$ and $N\!M\!I$ across all benchmark networks. These findings further substantiate the effectiveness of TAS-Com in addressing the CD problem when compared to methods employing graph pooling techniques.

\begin{table*}[htb!]
   \centering
   \resizebox{.95\textwidth}{!}{
   \begin{tabular}{ccccccccccccc}
   \toprule
   Approach & \multicolumn{2}{c}{Cora} & \multicolumn{2}{c}{Citeseer} & \multicolumn{2}{c}{Amazon Photo} & \multicolumn{2}{c}{Amazon PC} & \multicolumn{2}{c}{Coauthor CS} & \multicolumn{2}{c}{Coauthor Phy}\\ 
   \midrule
    & $Q$ & $N\!M\!I$ & $Q$ & $N\!M\!I$ & $Q$ & $N\!M\!I$ & $Q$ & $N\!M\!I$ & $Q$ & $N\!M\!I$ & $Q$ & $N\!M\!I$ \\
    \midrule
   MinCutPool & 70.3 & 35.8 & 78.9 & 25.9 & N/C & N/C & N/C & N/C & 70.5 & 64.6 & 64.3 & 48.3 \\
   \midrule
   Ortho & 65.6 & 38.4 & 74.5 & 26.1 & N/C & N/C & N/C & N/C & 65.7 & 64.6 & 59.5 & 44.7 \\
   \midrule
   \textbf{TAS-Com} & \textbf{81.7} & \textbf{65.1} & \textbf{88.1} & \textbf{41.0} & \textbf{72.2} & \textbf{78.4} & \textbf{61.5} & \textbf{61.2} & \textbf{74.1} & \textbf{82.1} & \textbf{67.3} & \textbf{66.0} \\
   \bottomrule
   \end{tabular}}
   \caption{Performance comparison in terms of $Q$ and $N\!M\!I$ (results are multiplied by 100) between TAS-Com and state-of-the-art approaches. The best results of $Q$ and $N\!M\!I$ are bolded. N/C: Not Converged. }\label{tab:performance}
\end{table*}

\section{Parameter settings of TAS-Com}

In the experiments, TAS-Com employs a GCN architecture identical to that proposed in \cite{bhowmick2024dgcluster}. Specifically, Table \ref{tab:parameterSettings} reports the parameter settings that we follow in the experiments of TAS-Com. 

\begin{table}[h]
   \centering
   \begin{tabular}{cc}
   \toprule
   \textbf{Parameter} & \textbf{Setting} \\ 
   \midrule
    Dimension of $W^{(0)}$ & $T \times 256$ \\
    Dimension of $W^{(1)}$ & $256 \times 128$ \\
    Dimension of $W^{(2)}$ & $128 \times 64$ \\
    Dimension of $X^{(e)}$ & $n \times 64$ \\
    Optimizer & Adam \\
    Learning rate & 0.001 \\
    Number of epochs & 300 \\
   \bottomrule
   \end{tabular}
   \caption{Parameter settings of TAS-Com in the experiments.}\label{tab:parameterSettings}
\end{table}  

\begin{figure*}[h]
    \centering
    \includegraphics[width=1.0\linewidth]{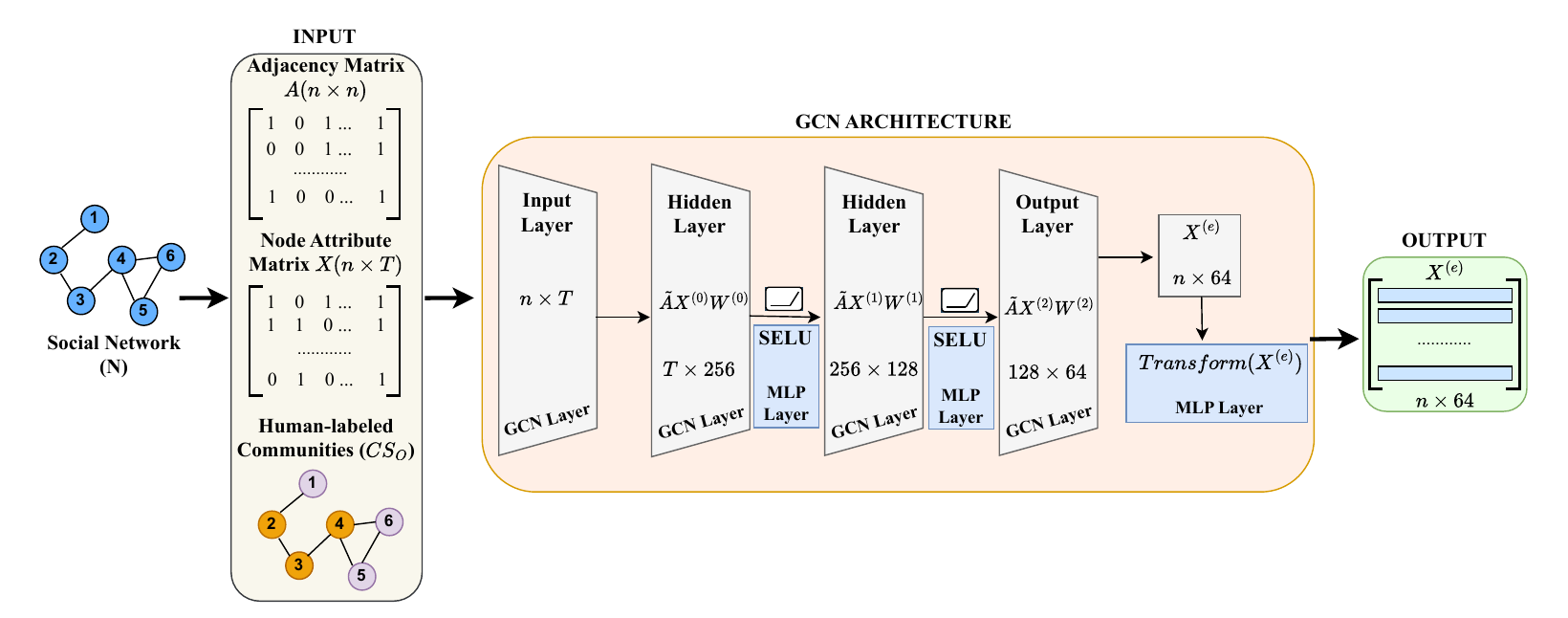}
    \caption{Graph Convolutional Neural Network architecture of TAS-Com.}
    \label{fig:GCNArchitecture}
\end{figure*}

Furthermore, Figure \ref{fig:GCNArchitecture} illustrates the architecture of the GCN as used in TAS-Com, which is designed to lean node embeddings by leveraging graph structure and node features. The input consists of an \textbf{adjacency matrix ($A$)}, which captures the connectivity of a given network $N$ ($A_{ij}$ = 1 if nodes $v_i$ and $v_j$ are connected, otherwise $A_{ij} = 0$), and a \textbf{node attribute matrix ($X$)}, where each row represents a node and each column corresponds to a feature, resulting in a matrix of size $n \times T$. Additionally, human-labeled communities $CS_O$ are provided as they are required to train the model. 

GCN comprises an \textbf{input layer}, \textbf{two hidden layers}, and an \textbf{output layer}. The input layer processes the feature matrix, $X$, using the normalized adjacency matrix, $\Tilde{A}$, and trainable weights, $W^{(0)}$, to generate the initial hidden representation. $\Tilde{A}$ is calculated as $\Tilde{A} = D^{-1/2}AD^{-1/2}$ with $D$ being the degree matrix of $A$. The first hidden layer transforms this input into a 256-dimensional representation, followed by the second hidden layer, which reduces it further to 128 dimensions, both employing the Scaled Exponential Linear Unit (SELU) activation function for non-linearity. The output layer produces the node embeddings with 64 dimensions, capturing rich node-level information. Then the obtained node embeddings from the output layer are further fine-tuned through a \textbf{Transformation} method using Equation \eqref{equ:L2Normalization} and then obtain the final node embedding, $X^{(e)}$. This architecture progressively aggregates and propagates information from neighboring nodes through $\Tilde{A}$, integrating both network topology and node features to extract meaningful representations.

\section{Definitions of the additional performance metrics}

To further evaluate the performance of the proposed TAS-Com, we employ two additional metrics: \emph{conductance}, a topology-based metric, and \emph{F1 score}, an attributive similarity-based metric \cite{bhowmick2024dgcluster}. 

\emph{\textbf{Conductance}} is a widely used metric for evaluating the quality of community structures in networks. It measures the fraction of edges that connect the nodes of a community to the rest of the graph compared to the total number of edges involving nodes within and outside the community. A community with low conductance indicates strong cohesion within the community and weak connections to the outside of the community, making it a good community by traditional definitions. Formally, the conductance $\phi(C_i)$ of a community $C_i$ in $CS$ is defined in Equation \eqref{equ:com_con}.

\begin{equation}\label{equ:com_con}
    \phi(C_i) = \frac{\text{Cut}(C_i,\overline{C_i})}{\text{min}(\text{Vol}(C_i), \text{Vol}(\overline{C_i}))},
\end{equation}

where $\text{Cut}(C_i, \overline{C_i})$ is the number of edges between the nodes in $C_i$ and the rest of the network $\overline{C_i}$. $\text{Vol}(C_i)$ is the sum of the degrees of all nodes in $C_i$. The denominator ensures normalization by considering the smaller volume of $C_i$ or $\overline{C_i}$. The above calculation of the conductance is performed for all the communities in the community structure $CS$ of a given network. Hence, the overall conductance $Con$ of a community structure $CS$ is defined as in Equation \eqref{equ:Con}.

\begin{equation}\label{equ:Con}
    Con =\frac{1}{|CS|} \sum\limits_{C_i \in CS}\phi(C_i).
\end{equation}

The \emph{\textbf{F1 score}} is a metric commonly used to evaluate CD algorithms by assessing the overlap between detected and ground truth communities. It is particularly useful when the task involves comparing the predicted community assignments with predefined community structures. The F1 score is the harmonic mean of \emph{precision} and \emph{recall}, capturing the trade-off between those two. Formally, F1 score $F1$ is defined as in Equation \eqref{equ:F1}.

\begin{equation}\label{equ:F1}
    F1 = 2 \cdot \frac{\text{Precision} \cdot \text{Recall}}{\text{Precision} + \text{Recall}}.
\end{equation}

In Equation \eqref{equ:F1}, \emph{Precision} represents the fraction of correctly identified community members among all those assigned to the community, whereas \emph{Recall} denotes the fraction of correctly identified members relative to the total number of true members in the ground truth community. A high F1 score (i.e., closer to 1) signifies a strong alignment between the detected community structure and the ground truth. Unlike conductance, which emphasizes structural cohesion, the F1 score evaluates the accuracy of node assignments based on predefined community labels, providing a complementary perspective for evaluating both topology and attribute-based community quality.

\section{Overview of the Leiden algorithm}
\begin{table*}[htb!]
   \centering
   \begin{tabular}{ccccccccc}
   \toprule
   Approach & \multicolumn{4}{c}{Cora} & \multicolumn{4}{c}{Amazon Photo} \\ 
   \midrule
    & $Q$ & $N\!M\!I$ & $Con$ & $F1$ & $Q$ & $N\!M\!I$ & $Con$ & $F1$ \\
    \midrule
    DGCluster($\lambda = 0.2$) & 80.77 & 46.93 & 11.86 & 33.19 & 72.67 & 65.99 & 12.98 & 59.35 \\
    DGCluster($\lambda = 0.8$) & 78.45 & \textbf{58.26} & 22.67 & \textbf{46.48} & 71.26 & \textbf{73.51} & 23.19 & \textbf{68.69} \\
    \midrule
    \textbf{TAS-Com} & \textbf{82.02} & 46.49 & \textbf{8.38} & 34.3 & \textbf{73.89} & 67.72 & \textbf{4.73} & 63.32 \\
   \bottomrule
   \end{tabular}
   \caption{Performance comparison of TAS-Com and DGCluster in terms of $Q$ (the higher the better), $N\!M\!I$ (the higher the better), $Con$ (the lower the better) and $F1$ (the higher the better) on unattributed networks. The best results (multiplied by 100) are bolded.}\label{tab:performanceUnattributed}
\end{table*}

The Leiden \cite{traag2019louvain} algorithm is an efficient algorithm for CD in unattributed networks, designed to address limitations of the Louvain \cite{blondel2008fast} algorithm, such as the formation of disconnected communities and suboptimal partitions. This algorithm aims to improve the quality and reliability of detected communities by ensuring that they are cohesive and structurally meaningful. The algorithm operates in three main phases: 1) the local moving phase, where nodes are moved between communities to optimize a quality function such as modularity while ensuring communities are connected; 2) the refinement phase, which splits communities into sub-communities if they contain disconnected components and refines them for better modularity; and 3) the aggregation phase, where communities are collapsed into super nodes for hierarchical analysis. These phases iterate until no further improvement is possible. 

Leiden algorithm has a complexity of $\mathrm{O}(m \log n)$, where n and m are the numbers of nodes and edges, respectively. Hence, the Leiden algorithm is suitable for analyzing large-scale networks in social networks, biological networks, and other domains, offering high-quality and more robust community structures than the Louvain algorithm.

\section{Performance evaluation of TAS-Com on unattributed networks}

We further evaluate the applicability of TAS-Com in the context of \emph{unattributed networks} by comparing its performance against our baseline algorithm, DGCluster. In scenarios where node attributes are unavailable, structural information—such as node adjacency, is commonly employed as a substitute for node features. Alternative structural representations, including node degree and Laplacian eigenvectors, may also serve this purpose in the absence of node attributes.

To simulate such conditions, we disregard the original node features of the Cora and Amazon Photo datasets, using instead the adjacency information to construct node representations. This approach is applied consistently across both TAS-Com and DGCluster to ensure a fair comparison. Table \ref{tab:performanceUnattributed} presents the performance comparison between TAS-Com and DGCluster in terms of four standard metrics: modularity ($Q$), normalized mutual information ($N\!M\!I$), conductance ($Con$), and F1-score ($F1$). As shown in the table, TAS-Com consistently outperforms DGCluster and its two variants with respect to $Q$ and $Con$—both of which are unsupervised evaluation metrics that rely solely on the network's connectivity structure. All experiments use the same value for the hyperparameter $\mu$, as detailed in Appendix B.1.

The superior performance of TAS-Com in terms of $Q$ and $Con$ demonstrates its effectiveness in detecting community structures based purely on topological information. However, TAS-Com does not achieve higher scores than DGCluster in $N\!M\!I$ and $F1$, which are metrics grounded in ground truth labels. Given that these labels in the benchmark datasets are derived from node attributes, their relevance diminishes when such attributes are intentionally omitted. Therefore, we argue that $N\!M\!I$ and $F1$ are less reliable indicators under the unattributed setting. Based on this analysis, we conclude that TAS-Com offers superior performance compared to state-of-the-art methods in unattributed network scenarios.

\section{Run time and memory usage analysis}
We assess TAS-Com's scalability by comparing its runtime and memory usage with our baseline algorithm, DGCluster, on both small (Cora) and large (Coauthor CS) networks. Table \ref{tab:runtime_analysis} and Table \ref{tab:memoryusage_analysis} report the analysis of runtime and memory usage analysis, respectively. According to the reported results, TAS-Com shows only marginal overhead compared to DGCluster. For example, on Cora, it takes on average 128s and 848MB compared to DGCluster’s 77s and 809MB. On Coauthor CS, TAS-Com takes 821s and 1.9GB versus DGCluster’s 734s and 1.7GB. The performance gap narrows on larger datasets, and memory usage remains stable. These results confirm that TAS-Com (using Leiden) offers comparable time and space complexity to DGCluster (without Leiden).

\begin{table}[H]
    \centering
    \begin{tabular}{ccc}
    \toprule
        Network & DGCluster & TAS-Com \\
    \midrule
        Cora & 77.23s & 128.41s\\
        Coauthor CS & 734.32s & 821.26s\\
    \bottomrule
    \end{tabular}
    \caption{Run time analysis of TAS-Com and DGCluster in terms of seconds (s).}
    \label{tab:runtime_analysis}
\end{table}

\begin{table}[H]
    \centering
    \begin{tabular}{ccc}
    \toprule
        Network & DGCluster & TAS-Com \\
    \midrule
        Cora & 809MB & 848MB\\
        Coauthor CS & 1.7GB & 1.9GB\\
    \bottomrule
    \end{tabular}
    \caption{Memory usage analysis of TAS-Com and DGCluster.}
    \label{tab:memoryusage_analysis}
\end{table}

\end{document}